\documentclass[letterpaper,twoside,journal]{IEEEtran}
\usepackage[latin9]{inputenc}
\usepackage{color}
\usepackage{verbatim}
\usepackage{float}
\usepackage{amsmath}
\usepackage{amsthm}
\usepackage{amssymb}
\usepackage{graphicx}
\usepackage{setspace}
\usepackage[hidelinks]{hyperref}

\makeatletter

\providecommand{\tabularnewline}{\\}
\floatstyle{ruled}
\newfloat{algorithm}{tbp}{loa}
\providecommand{\algorithmname}{Algorithm}
\floatname{algorithm}{\protect\algorithmname}

\let\oldforeign@language\foreign@language
\DeclareRobustCommand{\foreign@language}[1]{%
  \lowercase{\oldforeign@language{#1}}}
\theoremstyle{plain}
\newtheorem{thm}{\protect\theoremname}
\theoremstyle{plain}
\newtheorem{prop}[thm]{\protect\propositionname}
\theoremstyle{plain}
\newtheorem{cor}[thm]{\protect\corollaryname}

\usepackage{algorithm,algpseudocode}
\usepackage{cuted}
\usepackage{cite}

\allowdisplaybreaks

\makeatother

\providecommand{\corollaryname}{Corollary}
\providecommand{\propositionname}{Proposition}
\providecommand{\theoremname}{Theorem}

\begin{document}
\bstctlcite{reflib}
\markboth{Preprint: IEEE Transactions on Signal Processing, VOL. 69, 2021, PP. 5611-5626,  DOI: 10.1109/TSP.2021.3111705. }{}
\title{Tracking Cells and their Lineages via Labeled Random Finite Sets }
\author{Tran Thien Dat Nguyen, Ba-Ngu Vo, Ba-Tuong Vo, Du Yong Kim, and Yu
Suk Choi\thanks{T.T.D. Nguyen, B.-N. Vo and B.-T. Vo are with the School of Electrical
Engineering, Computing and Mathematical Sciences, Curtin University,
Australia (emails: t.nguyen172@postgrad.curtin.edu.au, \{ba-ngu.vo,
ba-tuong.vo\}@curtin.edu.au). D.Y. Kim is with the School of Engineering,
RMIT University, Australia (email: duyong.kim@rmit.edu.au). Y.S. Choi
is with the School of Human Sciences, University of Western Australia,
Australia (email: yusuk.choi@uwa.edu.au). \textit{Corresponding author:
T.T.D. Nguyen.} }\thanks{This work is supported by the Australian Research Council under Discovery
Project DP160104662, the Vice-Chancellor\textquoteright s Research
Fellowship, RMIT University, and the Heart Foundation Future Leader
Fellowship 101173.\textit{ }} \thanks{MATLAB implementation and associated datasets are available at \url{https://github.com/TranThienDat-Nguyen/RFS-cell-tracking}.} }
\maketitle
\begin{abstract}
Determining the trajectories of cells and their lineages or ancestries
in live-cell experiments are fundamental to the understanding of how
cells behave and divide. This paper proposes novel online algorithms
for jointly tracking and resolving lineages of an unknown and time-varying
number of cells from time-lapse video data. Our approach involves
modeling the cell ensemble as a labeled random finite set with labels
representing cell identities and lineages. A spawning model is developed
to take into account cell lineages and changes in cell appearance
prior to division. We then derive analytic filters to propagate multi-object
distributions that contain information on the current cell ensemble
including their lineages. We also develop numerical implementations
of the resulting multi-object filters. Experiments using simulation,
synthetic cell migration video, and real time-lapse sequence, are
presented to demonstrate the capability of the solutions.
\end{abstract}


\section{Introduction\label{sec:Introduction}}

Tracking cells from time-lapse video data is one of the foremost tasks
in developmental cell biology \cite{CTP036} and is critical to the
understanding of the laws governing cell behavior in living tissue\textendash laws
that predict when a cell will divide or differentiate into a specialized
cell type \cite{I003}. Cell tracking is a challenging problem due
to intricate cell motion/interaction, cell division/death, complex
sources of uncertainty, such as false positives, false negatives \cite{CTP005,CTP002,CTP001}.
A typical time-lapse video consists of thousands of frames. Thus manually
tracking the cells is time consuming and prone to human errors. Moreover,
given the large and growing volume of data, it becomes necessary to
automate cell tracking \cite{CTP045,CTP-B002,I004,CTP036,CTP002,CTP001,CTP031,CTP044,CTP022}. 

In addition to determining the trajectories of the cells, it is necessary
to resolve their lineages or ancestries in cell divisions as time
progresses. A cell\textquoteright s lineage describes the sequence
of ancestors of a cell and is important for the understanding of the
relationship between cell ancestry and cell fate\textendash how a
particular cell develops into a final cell type \cite{CTP046}. Many
solutions have been proposed for resolving cell lineages, see for
example \cite{CTP046} and references therein. However, most of these
methods are invasive in the sense that they require injection of dye
as markers to keep track of the cells and their ancestors. Non-invasive
solutions are more economically viable and suitable for almost all
types of experiments.\textcolor{blue}{{} }

Regardless of whether cell tracking is performed manually or automated,
it is important to note that the tracking results are not perfect.
Further, since cell experiments are mostly designed to infer certain
variables/parameters from the estimated tracks, errors in the inferred
results inevitable. How meaningful are the observations from the experiments
depend on the level of confidence in the inferred results. Hence,
it is important that the tracking framework has the capability to
characterize confidence on the inferred information.

So far, amongst the many approaches to multi-object tracking, Mahler's
random finite set (RFS) framework \cite{B001,B006} has a demonstrable
capability for characterizing confidence/uncertainty on the inferred
variables/results \cite{038}. Classical probabilistic multi-object
tracking approaches such as Multiple Hypothesis Tracking (MHT) \cite{054}
and Joint Probabilistic Data Association (JPDA) \cite{021}, have
been used in many applications, including cell tracking. However,
while they provide some form of confidence on the estimated tracks
individually, the issue of confidence on the variables inferred from
the tracks have not been considered. The RFS approach has also been
applied to cell tracking in \cite{CTP005,CTP001,CTP003,CTP047,CTP022}.
In \cite{026} a labeled RFS multi-object tracking filter that accommodates
spawning was proposed. However, such spawning model does not capture
ancestry in cell division, nor changes in cell appearance before dividing,
hence heuristic post-processing is needed to accommodate cell ancestry
\cite{CTP047}.

In this work, we propose a tractable spawning model and multi-object
tracking filters that address cell division including lineages and
changes in appearance. The labeled RFS formulation \cite{005} enables
lineage information to be encoded into the labels of individual objects.
Furthermore, the labels that identify individual objects and their
ancestry, can be naturally assimilated into the RFS spawning models,
and subsequently inferred from the data using labeled RFS estimation
techniques. The salient features of the proposed spawning model is
its ability to capture changes in cell appearance prior to division
and cell ancestry, thereby enabling better lineage estimation. When
a track is born from spontaneous birth, its label contains information
pertaining to when it is born and from which birth region \cite{005}.
Similarly, for a spawned track, its label contains information pertaining
to when and from which parent it originated. Under the proposed spawning
model, we derive the optimal multi-object Bayes tracking filter, and
two approximate filters, using moment and cardinality matching \cite{032},
as trade-offs between accuracy and computational load. Efficient implementations
are developed to operate under real world conditions where the (time-varying)
clutter rate and detection probability are not known. We also demonstrate
the capability of our approach to quantify confidence in the inferred
results.

For the remainder of the paper, we summarize related works and the
RFS framework in Section \ref{sec:Background}. Section \ref{sec:Labeled-RFS-tracker}
presents the novel spawning model and corresponding multi-object tracking
filters. Section \ref{sec:Cell-Tracking-Filter} details the implementations
of the cell tracking filters, and numerical studies are presented
in Section \ref{sec:Experimental-results}. 

\section{Background\label{sec:Background}}

\subsection{Related Work\label{subsec:Related-Work}}

There are two main approaches to cell tracking, namely model evolution
and track-by-detection. In model evolution, segmentation and tracking
(including the deformation of shapes) are carried out simultaneously.
On the other hand, track-by-detection treats detection and tracking
as two separate modules. 

\begin{table}[t]
\begin{onehalfspace}
\caption{List of symbols}

\end{onehalfspace}
\centering{}%
\begin{tabular}{|l|l|}
\hline 
{\small{}\,\,\,\,\,\,\,\,Notation} & {\small{}\quad{}\quad{}\quad{}\quad{}\quad{}\quad{}\,\,Description}\tabularnewline
\hline 
{\small{}$\mathbb{X}$} & {\small{}single object state space}\tabularnewline
{\small{}$\mathbb{L}$} & {\small{}label space}\tabularnewline
{\small{}$\mathbb{M}$} & {\small{}mode space}\tabularnewline
{\small{}$\mathbb{K}$} & {\small{}kinematic/feature space}\tabularnewline
{\small{}$\boldsymbol{X}$} & {\small{}labeled multi-object state}\tabularnewline
{\small{}$\boldsymbol{x}=(x,\ell)$} & {\small{}labeled single-object state (with label $\ell$)}\tabularnewline
{\small{}$m$} & {\small{}object mode}\tabularnewline
{\small{}$\zeta$} & {\small{}object kinematic state}\tabularnewline
{\small{}$[h(\cdot)]^{X}$} & {\small{}set exponential}\tabularnewline
{\small{}$\langle f,g\rangle$} & {\small{}inner product between $f$ and $g$}\tabularnewline
{\small{}$\delta_{Y}[X]$} & {\small{}generalized Dirac delta function}\tabularnewline
{\small{}$1_{Y}(X)$} & {\small{}set inclusion function}\tabularnewline
{\small{}$\left\langle \boldsymbol{f}\right\rangle $} & {\small{}label marginal of $\boldsymbol{f}$}\tabularnewline
{\small{}$\mathcal{L}\left(\boldsymbol{X}\right)$} & {\small{}set of labels of $\boldsymbol{X}$}\tabularnewline
{\small{}$\mathcal{F}(X)$} & {\small{}class of finite subsets of $X$}\tabularnewline
{\small{}$C$} & {\small{}maximum cardinality of generated sets }\tabularnewline
{\small{}$c$} & {\small{}cardinality of a generated set}\tabularnewline
{\small{}$\mathbb{G}_{+}^{(c)}(\ell)$} & {\small{}labels of set with cardinality $c$ generated by }\tabularnewline
 & {\small{}object labeled $\ell$}\tabularnewline
{\small{}$\mathbb{G}_{+}(\ell)$} & {\small{}labels of all sets generated by object labeled $\ell\!\!$ }\tabularnewline
{\small{}$\Phi_{+}^{(c)\!}(\cdot|\boldsymbol{x})$} & {\small{}joint labeled state density of $c$ daughters of $\boldsymbol{x}$}\tabularnewline
{\small{}$\varphi_{+}^{(c)}\left(\cdot\mid\zeta,\ell\right)$} & {\small{}joint kinematics-label density of $c$ daughters}\tabularnewline
 & {\small{}of $(\zeta,\ell)$}\tabularnewline
{\small{}$\vartheta^{(c)}(m_{+}^{(i)}|m,\ell)\!\!$} & {\small{}mode transition probability}\tabularnewline
{\small{}$\boldsymbol{g}(Z|\boldsymbol{X})$} & {\small{}multi-object likelihood}\tabularnewline
{\small{}$\varrho$} & {\small{}observed kinematic feature}\tabularnewline
{\small{}$\alpha$} & {\small{}observed appearance feature}\tabularnewline
{\small{}$\gamma$} & {\small{}extended association map}\tabularnewline
{\small{}$\Gamma$} & {\small{}space of (extended) association maps}\tabularnewline
{\small{}$\mathbb{D}$} & {\small{}space of association maps with division}\tabularnewline
{\small{}$\mathbb{N}$} & {\small{}space of association maps without division}\tabularnewline
\hline 
\end{tabular}
\end{table}
Algorithms in the model evolution category are usually based on minimizing
some energy functions via active contours \cite{CTP037,CTP020}, level
sets \cite{CTP025,CTP027,CTP-D012}, or mean shift \cite{CTP038}.
While this approach is accurate in tracking the cell membranes, there
are a number of disadvantages. Firstly, it is domain-specific as the
modeling of the contour evolution depends on the types of cell. Secondly,
it is computational intensive in high cell density scenarios \cite{CTP034},
which limits application to large scale problems. Thirdly, since the
detection and tracking modules cannot be separated, these algorithms
are not flexible and their performances degrade when the sampling
rate is low as the deformation cannot be adequately tracked \cite{CTP031}.

Track-by-detection infers cell tracks from the detector output, which
allows the extraction of temporal information on the cell population,
and the applications of different detection methods without reformulating
the tracking module. Track-by-detection algorithms can be further
classified as deterministic or probabilistic. Deterministic algorithms
are based on deterministically matching detections to cell tracks
by optimizing some cost functions \cite{CTP036,CTP014a,CTP-D001,CTP-D010},
and perform relatively well in scenarios where the cells are well
separated. However, performance deteriorates when the clutter rate
and cell density are high \cite{CTP002}. In probabilistic algorithms,
the cell tracks are inferred from some form of probability distribution
\cite{CTP-B002,CTP001,CTP002}. This approach has been demonstrated
to track closely spaced cells in environments with high clutter rate
and low detection probability \cite{CTP002}.

The quality of cell detectors influences the performance of cell trackers
in the track-by-detection approach. Apart from segmentation in the
model evolution approach, there are generally two approaches to detect
cells: morphological thresholding and machine learning. In morphological
thresholding, image filtering is applied to remove noise, followed
by locally adaptive thresholding \cite{D010,D004,D014,D011}, and
finally, size filtering to obtain cell blobs. With the advent of neural
networks, machine learning for cell detection is gaining attention
\cite{D013,D012,D009}. Detectors in this category are usually neural
networks trained to predict enclosed boxes, segmentation masks and
location of cells from input images. While promising, high computational
efforts and large volumes of training data are needed to achieve reliable
predictions. In practice, the morphological approach is still widely
used given its speed and relative accuracy. Moreover, in fluorescence
imaging, the cells are usually observed as bright/dark spots without
any features, and hence rely mainly on morphological thresholding
for detection.

Due to its importance in developmental cell biology, a number of cell
lineage estimation techniques have been developed \cite{CTP030}.
The tracking-free methods rely on features associated with mitotic
cells to detect mitosis in individual images \cite{CTP040,CTP041,CTP042}.
The tracking-based methods identify mitosis by integrating mitotic
models/classifiers into the cell trackers\textcolor{blue}{{} }\cite{CTP-D010,CTP-D016,HKBCK2011mitodet,CTP038,CR2015contawarecell,XLSCN2021antcolony},
or by post-processing the tracking results to construct the lineage
tree \cite{CTP039}. While tracking-free methods can perform well
in high mitosis abnormality levels, tracking-based methods are more
advantageous when this abnormality level is weak or the image sampling
rate is low.

Established approaches such as JPDA \cite{021}, MHT \cite{054} and
RFS \cite{B001,B006} have been applied to cell tracking in \cite{CTP005,CTP002,066,CTP001,CTP011,CTP047}.
The RFS approach models the entire ensemble of cells as an RFS, which
naturally encapsulates the uncertainty in the cell population due
to mitosis, migration, death, and the presence of clutter and mis-detection.
More importantly, labeled RFS provides natural means for modeling
cell trajectories and their lineages \cite{038}, \cite{026}. Numerically,
the RFS approach has been demonstrated on very large-scale problems
\cite{040}, and hence promising for applications with high cell density.

For most multi-object trackers, detection probability and clutter
rate are important prior parameters that are usually assumed known.
However, in biological applications, these parameters are unknown
and vary with time. RFS-based filters have been developed to address
this problem in \cite{009,043,065}. In \cite{CTP001}, the robust
CPHD filter \cite{009} that estimates clutter rate and detection
probability, was bootstrapped to another standard CPHD filter \cite{003}
to track cells. However, this algorithm does not consider cell division.
Moreover, the robust CPHD filter \cite{009} is superseded by the
newer solution in \cite{043}.

\subsection{Bayesian Multi-Object Filtering \label{subsec:Bayes-Filter}}

In the (classical) Bayes filter, all information on the current state
$x$, modeled as a random vector, is encapsulated in the filtering
density $p$ (which is conditioned on the observation history, but
omitted for clarity). Moreover, this density can be propagated to
the next time via the Bayes recursion \cite{B005}
\begin{eqnarray}
p_{+}\!\left(x_{+}\right) & = & {\scriptstyle {\displaystyle {\textstyle \int}}}f_{+}\!\left(x_{+}|x\right)p\left(x\right)dx,\\
p_{+}\!\left(x_{+}|z_{+}\right) & \propto & g_{+}\left(z_{+}|x_{+}\right)p_{+}\!\left(x_{+}\right),
\end{eqnarray}
where $p_{+}\!$ is the prediction density, $f_{+}(x_{+}|x)$ is the
Markov transition density to the state $x_{+}$ from a given $x$,
and $g_{+}(z_{+}|x_{+})$ is the likelihood that $x_{+}$ generates
an observation $z_{+}$. For simplicity we omit the subscript for
current time and use the subscript `+' to denote the next time step. 

Cell tracking is a multi-object estimation problem because the number
of cells and their states are unknown and time-varying. Thus, instead
of a single state vector we have a set of state vectors, called the
\textit{multi-object state}. Specifically, each element of the multi-object
state $\boldsymbol{X}$ is an ordered pair $\boldsymbol{x}=(x,\ell)$,
where $x$ is a state vector in some space $\mathbb{X}$, and $\ell$
is a distinct label in some discrete space $\mathbb{L}$ \cite{005}.
The label of an $\boldsymbol{x\in}\mathbb{X}\times\mathbb{L}$ is
given by the \textit{label extraction function} $\mathcal{L}\left(\boldsymbol{x}\right)$,
and the labels of any $\boldsymbol{X\subset}\mathbb{X}\times\mathbb{L}$
is defined as $\mathcal{L}\left(\boldsymbol{X}\right)\triangleq\left\{ \mathcal{L}\left(\boldsymbol{x}\right):\boldsymbol{x}\in\boldsymbol{X}\right\} $.

Hereon, we adhere to the following notations: 
\begin{multline*}
\begin{array}{cc}
[h(\cdot)]^{X}\triangleq\prod_{x\in X}h(x); & \ \ \langle f,g\rangle\triangleq\int f(x)g(x)dx;\\
\delta_{Y}[X]\triangleq\begin{cases}
1 & X=Y\\
0 & X\neq Y
\end{cases}; & \ \ 1_{Y}(X)\triangleq\begin{cases}
1 & X\subseteq Y\\
0 & \textrm{otherwise}
\end{cases}.
\end{array}
\end{multline*}
For a singleton $X=\{x\}$, we abbreviate $1_{Y}(x)\triangleq1_{Y}(\{x\})$.
Since the multi-object state $\boldsymbol{X}$ must have distinct
labels, we require the \emph{distinct label indicator} $\Delta\left(\boldsymbol{X}\right)\triangleq\delta_{\left|\boldsymbol{X}\right|}\left[\left|\mathcal{L}\left(\boldsymbol{X}\right)\right|\right]$
be equal to 1. We also denote the class of finite subsets of a space
$S$ by $\mathcal{F}(S)$, and for a function $f:\mathcal{F}(\mathbb{X}\times\mathbb{L})\rightarrow\mathbb{R}$,
we define its \textit{label-marginal} $\left\langle f\right\rangle :\mathcal{F}(\mathbb{L})\rightarrow\mathbb{R}$,
by
\begin{multline}
\left\langle f\right\rangle (\{\ell_{1},...,\ell_{n}\})\triangleq{\textstyle \int}f(\{(x_{1},\ell_{1}),...,(x_{n},\ell_{n})\})dx_{1:n}.
\end{multline}

In line with the Bayesian paradigm, the multi-object state is modeled
as a random finite set, characterized by Mahler's multi-object density
\cite{B001,B006} (equivalent to a probability density \cite{001}).
The multi-object Bayes filter takes on the same form as the classical
Bayes filter with: $x$ and $x_{+}$ replaced by the sets $\boldsymbol{X}$
and $\boldsymbol{X}_{+}$ of multi-object states; $p$, $p_{+}$ and
$p_{+}(\cdot|z_{+})$ replaced by the multi-object filtering/prediction
densities $\mathbf{\boldsymbol{\pi}}$, $\mathbf{\boldsymbol{\pi}_{+}\!}$
and $\boldsymbol{\pi}_{+}(\cdot|Z_{+})$; $f_{+}$ and $g_{+}$ replaced
by the multi-object transition density $\boldsymbol{f}_{+}$ and multi-object
(observation) likelihood $\boldsymbol{g}_{+}$; $z_{+}$ replaced
by the measurement set $Z_{+}$; and the vector integral replaced
by the set integral \cite{B006}, i.e.
\begin{eqnarray}
\boldsymbol{\pi}_{+}\!\left(\boldsymbol{X}_{+}\right) & = & {\textstyle \int}\boldsymbol{f}_{+}\!\left(\boldsymbol{X}_{+}|\boldsymbol{X}\right)\boldsymbol{\pi}\left(\boldsymbol{X}\right)\delta\boldsymbol{X},\label{e:MTBF}\\
\boldsymbol{\pi}_{+}\!\left(\boldsymbol{X}_{+}|Z_{+}\right) & \propto & \boldsymbol{g}_{+}\left(Z_{+}|\boldsymbol{\boldsymbol{X}}_{+}\right)\boldsymbol{\pi}_{+}\!\left(\boldsymbol{X}_{+}\right).\label{e:MTBF-1}
\end{eqnarray}

The multi-object (observation) likelihood captures the observation
noise, false negatives, and false positives. For $Z=\{z_{1:|Z|}\}$,
the multi-object likelihood is given by
\begin{equation}
\boldsymbol{g}(Z|\boldsymbol{X})\propto\sum_{\theta\in\Theta}1_{\Theta(\mathcal{L}(\boldsymbol{X}))}(\theta)\left[\varPsi_{Z}^{(\theta)}\right]^{\boldsymbol{X}},\label{eq:multi_likelihood}
\end{equation}
where $\Theta$ is the set of positive 1-1 maps $\theta$ taking the
object labels to indices of observations, $\Theta\left(I\right)$
is the subset of $\Theta$ with domain $I$, $\varPsi_{Z}^{(\theta)}(x,\ell)=\psi_{Z}^{(\theta(\ell))}(x,\ell)$,
\[
\psi_{Z}^{(j)}(x,\ell)=\begin{cases}
\frac{P_{D}(x,\ell)g(z_{j}|x,\ell)}{\kappa(z_{j})} & \textrm{if }j\in\{1,...,|Z|\}\\
1-P_{D}(x,\ell) & \textrm{if }j=0
\end{cases},
\]
$\kappa(\cdot)$ is the clutter intensity, $P_{D}(x,\ell)$ is the
detection probability, and $g(\cdot|x,\ell)$ is the single-object
likelihood function \cite{006}. 

The multi-object transition density captures the motions, births and
deaths of objects. Births can occur independently or spawn from parent
objects. A multi-object transition density that models spawning was
proposed in \cite{026}. However, this model does not accommodate
ancestry in cell division because it assumes independence between
the parent's existence and spawned objects' existence. Specifically,
a parent can survive/die independent of whether it spawns or not.
This is not the case in cell division where the parent ceases to exist
at the moment it spawns. Hence, the model in \cite{026} cannot capture
mitosis because it permits parents and daughters to co-exist. 

\section{Labeled RFS Tracker for Cell Biology\label{sec:Labeled-RFS-tracker}}

This section presents a multi-object transition density that captures
cell lineages (Subsection \ref{subsec:Multi-object-models}), the
resulting multi-object filters (Subsection \ref{subsec:Multi-object-filtering}),
and a cell division model that incorporates cell appearance (Subsection
\ref{subsec:Cell-appearance}). Extension to tracking with unknown
clutter rate, detection probability, and birth parameters is discussed
in Subsection \ref{subsec:UnknownBG}.

\subsection{Spawning Model for Cell Division\label{subsec:Multi-object-models}}

Following \cite{005}, the label $\ell=(k+1,\iota)$ of a spontaneous
birth, i.e. a new object with no parent, comprises the time of birth,
and an index to distinguish those born at the same time. Hence, the
space $\mathbb{B}_{+}$ of spontaneous birth labels at time $k+1$
is $\{k+1\}\times\mathbb{I}$, where $\mathbb{I}$ is a discrete set. 

\begin{figure}
\begin{centering}
\includegraphics[width=0.45\textwidth]{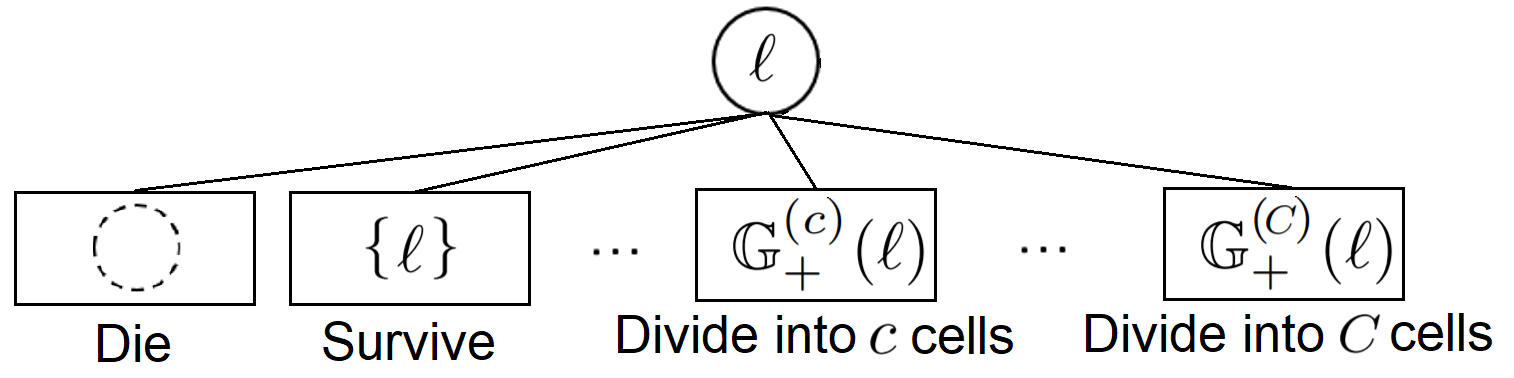}
\par\end{centering}
\caption{An object can divide into maximum of $C$ objects in the next time.\label{fig:cell-div-model}}
\end{figure}

An object (with label $\ell$) can generate, at the next time, a set
$\mathbb{G}_{+}^{(c)}(\ell)$ of $c$ objects with \textit{distinct}
labels. For: $c=0$, the object dies and $\mathbb{G}_{+}^{(0)}(\ell)=\emptyset$;
$c=1$, the object continues to live and $\mathbb{G}_{+}^{(1)}(\ell)=\{\ell\}$;
and $c>1$, the object spawns $c$ daughters with label set $\mathbb{G}_{+}^{(c)}(\ell)=\{(\ell,k+1,c)\}\times\{1,...,c\}$,
see Fig. \ref{fig:cell-div-model}. In this convention the label $(\ell,k+1,c,\iota)\in\mathbb{G}_{+}^{(c)}(\ell)$
consists of the parent label, the time of birth, the number of siblings,
and an index to distinguish it amongst the siblings. Assuming $\ell$
can spawn at most $C$ daughters at a time (for cell division $C=2$
since a cell can only divide into two), the space of possible labels
generated by $\ell$ is $\mathbb{G}_{+}(\ell)\triangleq\biguplus_{c=1}^{C}\mathbb{G}_{+}^{(c)}(\ell)$.
We also abbreviate $\mathbb{G}_{+}(\boldsymbol{x})=\mathbb{G}_{+}(\mathcal{L}(\boldsymbol{x}))$,
and $\mathbb{G}_{+}(L)\triangleq\biguplus_{\ell\in L}\mathbb{G}_{+}(\ell)$.

Given the current label space $\mathbb{L}$, the space of all possible
labels at the next time is $\mathbb{L}_{+}=\mathbb{G}_{+}(\mathbb{L})\uplus\mathbb{B}_{+}$.
Note that $\mathbb{L}=\biguplus_{\ell\in\mathbb{L}}\mathbb{G}_{+}^{(1)}(\ell)\subset\mathbb{L}_{+}$.
To address lineage, we define the Parent function on $\mathbb{L}_{+}-\mathbb{B}_{+}-\mathbb{L}$
by $\text{Parent}((\ell,k+1,c,\iota))=\ell$. 

\textit{Remark}. There are no unique labeling conventions for spawned
objects. For consistency in lineage, we require the following conditions
(which our labeling convention satisfies):
\begin{enumerate}
\item $\mathbb{G}_{+}^{(c)}(\ell)\cap\mathbb{G}_{+}^{(c')}(\ell')=\emptyset$
if $\ell\neq\ell'$, i.e. the children from two distinct objects must
not share any common labels; 
\item $\mathbb{G}_{+}^{(c)}(\ell)\cap\mathbb{G}_{+}^{(c')}(\ell)=\emptyset$
if $c\neq c'$, i.e. any two sets of children with different cardinalities,
(even from the same parent), should not have common labels.
\end{enumerate}
These properties enable the parent of any $\mathbb{\ell\in\mathbb{L}_{+}-\mathbb{B}_{+}-\mathbb{L}}$
to be determined as the (unique) label $\ell'$ such that $\ell\in\mathbb{G}_{+}(\ell')$. 

The new set of objects generated from a single object with labeled
state $\boldsymbol{x}$ is modeled by a labeled RFS with density:
\begin{eqnarray}
\!\!\!\boldsymbol{f}_{\!G,+}(\boldsymbol{U}|\boldsymbol{x})=\Delta(\boldsymbol{U})\sum_{c=0}^{C}\delta_{\mathbb{G}_{+}^{(c)\!}(\boldsymbol{x})}[\mathcal{L}(\boldsymbol{U})]\rho_{+}^{(c)}(\boldsymbol{x})\Phi_{+}^{(c)}(\boldsymbol{U}|\boldsymbol{x}),\!\!\!\!\label{eq:Transition-with-CD}
\end{eqnarray}
where $\rho_{+}^{(c)}(\boldsymbol{x})$ is the probability that $\boldsymbol{x}$
generates $c$ objects at the next time, and $\Phi_{+}^{(c)}(\{\boldsymbol{x}_{+}^{(1)},...,\boldsymbol{x}_{+}^{(c)}\}|\boldsymbol{x})$
is the joint density of their $c$ states, with the convention $\Phi_{+}^{(0)}(\emptyset|\boldsymbol{x})=1$.
Note that parent objects cannot co-exist with their daughters, and
the kinematics/features of siblings from the same parent are statistically
dependent. On the other hand, in \cite{026} parents of spawned objects
can continue to exist, and the kinematics/features of siblings from
the same parent are statistically independent.

It is assumed that the sets of objects generated by individual elements
of a given (labeled) multi-object state $\boldsymbol{X}$ are independent
of each other and the set of spontaneous births. Additionally, since
all generated labels are distinct, the multi-object state $\boldsymbol{X}_{+}$
at the next time is a disjoint union of the spontaneous births and
the sets of labeled states generated from different elements of $\boldsymbol{X}$.
Hence, using the FISST convolution theorem \cite{B001} the multi-object
transition density is given by
\begin{multline}
\!\!\boldsymbol{f}_{+}(\boldsymbol{X}_{+}\mid\boldsymbol{X})=\boldsymbol{f}_{B,+}(\boldsymbol{X}_{+}\cap(\mathbb{X}\times\mathbb{B}_{+}))\boldsymbol{f}_{G,+}(\boldsymbol{X}_{+}|\boldsymbol{X}),\label{eq:exact_transition}
\end{multline}
where $\boldsymbol{f}_{B,+}$ is the density of the spontaneous-birth
set, and 
\begin{multline}
\!\boldsymbol{f}_{G,+}(\boldsymbol{X}_{+}|\boldsymbol{X})=\prod_{\boldsymbol{x}\in\boldsymbol{X}}\boldsymbol{f}{}_{G,+}(\boldsymbol{X}_{+}\cap(\mathbb{X}\times\mathbb{G}_{+}(\boldsymbol{x}))|\boldsymbol{x}).\!\!
\end{multline}
A popular birth model is a labeled multi-Bernoulli (LMB) \cite{005}
\begin{eqnarray}
\boldsymbol{f}_{B,+}(\boldsymbol{Y}) & = & \Delta(\boldsymbol{Y})w_{B,+}(\mathcal{L}(\boldsymbol{Y}))[p_{B,+}]^{\boldsymbol{Y}},\label{eq:LMB_birth}
\end{eqnarray}
where $w_{B,+}(L)=1_{\mathbb{B}_{+}}(L)[1-r_{B,+}]^{\mathbb{B}_{+}-L}[r_{B,+}]^{L},$
$r_{B,+}(\ell)$ is the probability of new a birth with label $\ell$,
and $p_{B,+}(\cdot,\ell)$ is the probability density of its (unlabeled)
state.

\textit{Remark.} In addition to the above labeling scheme, our proposed
dynamic model removes the independence in the parent's existence and
daughters' existence assumed in \cite{026}. Consequently, the techniques
for propagating the multi-object filtering density used in \cite{026}
is no longer applicable.

\subsection{Multi-object Filtering with Cell Division\label{subsec:Multi-object-filtering}}

For the standard multi-object system model (no spawnings), if the
initial prior is a Generalized Labeled Multi-Bernoulli (GLMB), then
the prediction and filtering densities are also GLMBs, i.e. multi-object
densities of the form \cite{005}
\begin{equation}
\boldsymbol{\pi}\left(\boldsymbol{X}\right)=\Delta\left(\boldsymbol{X}\right)\sum_{I,\xi}\omega^{\left(I,\xi\right)}\delta_{I}[\mathcal{L}(\boldsymbol{X})]\left[p^{(\xi)}\right]^{\boldsymbol{X}},\label{e:GLMB-1}
\end{equation}
where $I\in\mathcal{F}(\mathbb{L})$, $\xi\in\Xi$ the space of all
association histories, each $\omega^{\left(I,\xi\right)}$ is non-negative
such that $\sum_{I,\xi}\!\omega{}^{(I,\xi)}=1$, and each $p^{\left(\xi\right)}\left(\cdot,\ell\right)$
is a probability density on $\mathbb{X}$. 

Hereon, we use the multi-object system model described by our proposed
multi-object transition density (\ref{eq:exact_transition}) and multi-object
likelihood (\ref{eq:multi_likelihood}). In this case the prediction
and filtering densities are not GLMBs, but take on a more general
form:
\begin{equation}
\boldsymbol{\pi}(\boldsymbol{X})=\Delta(\boldsymbol{X})\!\sum_{I,\xi}\!\omega^{(I,\xi)}\delta_{I}[\mathcal{L}(\boldsymbol{X})]p^{(\xi)}(\boldsymbol{X}),\label{eq:general_init}
\end{equation}
where $\left\langle p^{(\xi)}\right\rangle (L)=1$, for each $L\in\mathcal{F}(\mathbb{L})$. 

The RFS framework provides the tools for characterizing uncertainty
in the ensemble of trajectories such as the cardinality distribution,
intensity %
\begin{comment}
(or Probability Hypothesis Density) 
\end{comment}
$v$, and some other statistics: 
\begin{gather}
\Pr(\left\vert \boldsymbol{X}\right\vert \text{=}n)=\sum_{I,\xi}\delta_{n}\left[\left\vert I\right\vert \right]\omega^{(I,\xi)},\label{eq:comp_stats_01}\\
\!\!\!\!\!\!\!v(x,\ell)=\sum_{I,\xi}1_{I}(\ell)\omega^{(I,\xi)}\left\langle p^{(\xi)\!}\left(\{(x,\ell)\}\!\uplus\!(\cdot)\right)\!\right\rangle \!(I\!-\!\{\ell\}),\\
\!\Pr(\text{label \ensuremath{\ell\ }exists})=\sum_{I,\xi}1_{I}(\ell)\omega^{(I,\xi)},\\
\!\!\!\!\!\Pr(n\text{ new spawnings})=\sum_{I,\xi}\delta_{n\!}\left[\left\vert I\mathbf{\cap}(\mathbb{L}\!-\!\mathbb{L}_{-}\!-\!\mathbb{B})\right\vert \right]\omega^{(I,\xi)},\\
\!\!\!\!\!\Pr(n\text{ divisions})\!=\!\sum_{I,\xi}\delta_{n\!}\left[\left\vert \text{Parent}\!\left(I\mathbf{\cap}(\mathbb{L}\!-\!\mathbb{L}_{-}\!-\!\mathbb{B})\right)\right\vert \right]\omega^{(I,\xi)}\!.\label{eq:comp_stats_05}
\end{gather}

From (\ref{eq:general_init}), trajectories are estimated by first
finding a cardinality $n^{*}$ that maximizes the cardinality distribution,
and the component $(I^{*},\xi^{*})$ with highest weight such that
$|I^{*}|=n^{*}$. The trajectories with labels in $I^{*}$ are estimated
jointly from the function $p^{(\xi^{*})}(\cdot)$. For the GLMB special
case, the trajectory of each $\ell\in I^{*}$ is estimated from $p^{(\xi^{*})}(\cdot,\ell)$
\cite{038,053}.

The propagation of the multi-object prediction and filtering densities
are given in Propositions \ref{prop:exact_prediction} and \ref{prop:exact_update},
respectively (see Appendix \ref{subsec:Proof-of-pred-aprx-1}, \ref{subsec:Proof-of-upd-aprx}
for proof).
\begin{prop}
\label{prop:exact_prediction}Given a current multi-object filtering
density of the form (\ref{eq:general_init}), the prediction density
at the next time is
\begin{multline}
\!\!\!\!\!\boldsymbol{\pi}_{\!+}\!(\boldsymbol{X}_{\!+})\!=\!\Delta(\boldsymbol{X}_{\!+})\!\!\sum_{I,\xi,I_{+}}\!\!\omega_{+}^{(I,\xi,I_{+\!})}\delta_{I_{+\!}}[\mathcal{L}(\boldsymbol{X}_{\!+})]p_{+}^{(I,\xi)}(\boldsymbol{X}_{\!+}),\label{eq:exact_prediction}
\end{multline}
where $I\subseteq\mathbb{L}$, $\xi\in\Xi$, $I_{+}\subseteq\mathbb{L}_{+}$,
and
\begin{eqnarray}
\!\!\!\!\!\!\!\!\!\omega_{+}^{(I,\xi,I_{+})} & \!\!\!= & \!\!\!\omega^{(I,\xi)}w_{B,+}(I_{+}\cap\mathbb{B}_{+})\eta_{G,+}^{(I,\xi)}(I_{+}-\mathbb{B}_{+}),\\
\!\!\!\!\!\!\!\!\!p_{+}^{(I,\xi)\!}(\boldsymbol{Y}) & \!\!\!= & \!\!\!\left[p_{B,+}\right]{}^{\boldsymbol{Y}\cap(\mathbb{X}\times\mathbb{B}_{+})}p_{G,+}^{(I,\xi)}(\boldsymbol{Y}-(\mathbb{X}\times\mathbb{B}_{+})),\\
\!\!\!\!\!\!\!\!\!\eta_{G,+}^{(I,\xi)}(L) & \!\!\!= & \!\!\!\left\langle q_{G,+}^{(I,\xi)}\right\rangle (L),\label{eq:eta_G_exact_pred}\\
\!\!\!\!\!\!\!\!\!p_{G,+}^{(I,\xi)}(\boldsymbol{Y}) & \!\!\!= & \!\!\!q_{G,+}^{(I,\xi)}(\boldsymbol{Y})/\eta_{G,+}^{(I,\xi)}(\mathcal{L}(\boldsymbol{Y})),\\
\!\!\!\!\!\!\!\!\!q_{G,+}^{(I,\xi)}(\boldsymbol{Y}) & \!\!\!= & \!\!\!\left\langle p^{(\xi)}(\cdot)\boldsymbol{f}_{G,+}(\boldsymbol{Y}|\cdot)\right\rangle (I).
\end{eqnarray}
Moreover, if the current multi-object filtering density is a GLMB
of the form (\ref{e:GLMB-1}), then
\begin{align}
\!\!\!\!\!\!q_{G,+}^{(I,\xi)}(\boldsymbol{Y}) & =\left[{\textstyle \int}p^{(\xi)\!}(x,\cdot)\boldsymbol{f}_{\!G,+\!}(\boldsymbol{Y}\!\cap\!(\mathbb{X\!}\times\!\mathbb{G}_{+}(\cdot))|x,\cdot)dx\right]^{I}\!\!\!\!.\label{eq:PredofGLMB}
\end{align}
\end{prop}
Note that for component $(I,\xi,I_{+})$ of the multi-object prediction
density, its weight $\omega_{+}^{(I,\xi,I_{+})}$ is the product of
the previous weight, the predictive probability of the new birth label
set, and the predictive probability of the label set of generated
objects. Similarly, its density $p_{+}^{(I,\xi)\!}$ is the product
of the predictive density of new birth objects and the predictive
density of generated objects.
\begin{prop}
\label{prop:exact_update}Given a current multi-object filtering density
of the form (\ref{eq:general_init}), the filtering density at the
next time given the multi-object measurement $Z_{+}$ is 
\begin{multline}
\!\!\!\!\boldsymbol{\pi}_{\!+}(\boldsymbol{X}_{+}|Z_{+})\propto\\
\Delta(\boldsymbol{X}_{\!+})\!\!\sum_{I,\xi,I_{+},\theta_{+}}\!\omega_{Z_{+}}^{(I,\xi,I_{+},\theta_{+})\!}\delta_{I_{+\!}}[\mathcal{L}(\boldsymbol{X}_{\!+})]p_{Z_{+}}^{(I,\xi,\theta_{+})\!}(\boldsymbol{X}_{\!+}),\label{eq:exact_filtering}
\end{multline}
where $I\subseteq\mathbb{L}$, $\xi\in\Xi$, $I_{+}\subseteq\mathbb{L}_{+},\theta_{+}\in\Theta_{+}$,
\begin{eqnarray}
\!\!\!\!\!\!\!\!\!\omega{}_{Z_{+}}^{(I,\xi,I_{+},\theta_{+})} & \!\!\!\!\!= & \!\!\!\!\!\omega_{+}^{(I,\xi,I_{+})}1_{\Theta_{+}(I_{+})}(\theta_{+})\times\nonumber \\
\!\!\!\!\!\!\!\!\! & \!\!\!\!\! & \!\!\!\!\!\left[\eta_{B,Z_{+}}^{(\theta_{+})}\right]^{I_{+}\cap\mathbb{B}_{+}}\eta_{G,Z_{+}}^{(I,\xi,\theta_{+})}(I_{+}-\mathbb{B}_{+}),\label{eq:exact_upd_weight}\\
\!\!\!\!\!\!\!\!\!\eta_{B,Z_{+}}^{(\theta_{+})}(\ell) & \!\!\!\!\!= & \!\!\!\!\!\left\langle p_{B,+}(\cdot,\ell),\varPsi_{Z_{+}}^{(\theta_{+})}(\cdot,\ell)\right\rangle ,\\
\!\!\!\!\!\!\!\!\!\eta_{G,Z_{+}}^{(I,\xi,\theta_{+})}(L) & \!\!\!\!\!= & \!\!\!\!\!\left\langle p_{G,+}^{(I,\xi)}(\cdot)\left[\varPsi_{Z_{+}}^{(\theta_{+})}\right]{}^{(\cdot)}\right\rangle (L),\label{eq:etaG}\\
\!\!\!\!\!\!\!\!\!p_{Z_{+}}^{(I,\xi,\theta_{+})\!}(\boldsymbol{Y}) & \!\!\!\!\!= & \!\!\!\!\!\left[p_{B,Z_{+}}^{(\theta_{+})}\!\right]\!^{\!\boldsymbol{Y}\cap(\mathbb{X}\!\times\!\mathbb{B}_{+\!})}p_{G,Z_{+}}^{(I,\xi,\theta_{+})\!}(\boldsymbol{Y}\!-\!(\mathbb{X}\!\times\!\mathbb{B}_{+\!})),\!\!\\
\!\!\!\!\!\!\!\!\!p_{B,Z_{+}}^{(\theta_{+})}(x,\ell) & \!\!\!\!\!\propto & \!\!\!\!\!p_{B,+}(x,\ell)\varPsi_{Z_{+}}^{(\theta_{+})}(x,\ell),\\
\!\!\!\!\!\!\!\!\!p_{G,Z_{+}}^{(I,\xi,\theta_{+})}(\boldsymbol{Y}) & \!\!\!\!\!\propto & \!\!\!\!\!p_{G,+}^{(I,\xi)}(\boldsymbol{Y})\left[\varPsi_{Z_{+}}^{(\theta_{+})}\right]{}^{\boldsymbol{Y}}.\label{eq:pG}
\end{eqnarray}
\end{prop}
Assuming that component $(I,\xi,I_{+},\theta_{+})$ of the multi-object
filtering density has valid association map $\theta_{+}$, i.e. $1_{\Theta_{+}(I_{+})}(\theta_{+})=1$,
then its weight $\omega{}_{Z_{+}}^{(I,\xi,I_{+},\theta_{+})}$ is
the product of the predictive weight, the data-updated weight of the
new birth label set, and data-updated weight of the label set of generated
objects. Similarly, its density $p_{Z_{+}}^{(I,\xi,\theta_{+})\!}$
is the product of the data-updated new birth object density and the
data-updated density of generated objects.

Unlike the GLMB recursion, propagating the multi-object filtering
density (\ref{eq:exact_filtering}) is numerically intensive, due
to the growing number of high-dimensional densities over time. To
alleviate this problem, we present two GLMB approximation strategies
based on prediction and update approximations.

\subsubsection*{Prediction Approximation}

This strategy approximates the prediction density by a GLMB using
Proposition 2 of \cite{032}, and then performs an (exact) GLMB update
\cite{005} to yield a GLMB approximate filtering density, as summarized
in Corollary \ref{cor:pred_apprx}.
\begin{cor}
\label{cor:pred_apprx}A GLMB that matches the prediction density
(\ref{eq:exact_prediction}) in first moment and cardinality distribution
is given by\textup{
\begin{align}
\!\!\widehat{\boldsymbol{\pi}}_{+}(\boldsymbol{X}_{+})=\enskip\enskip\enskip\enskip\qquad\qquad\qquad\enskip\qquad\qquad\qquad\qquad\nonumber \\
\Delta(\boldsymbol{X}_{\!+})\!\!\sum_{I,\xi,I_{+}}\!\!\omega_{+}^{(I,\xi,I_{+})}\delta_{I_{+\!}}[\mathcal{L}(\boldsymbol{X}_{\!+})]\left[p_{+}^{(I,\xi,I_{+})}\right]^{\boldsymbol{X}_{\!+}}\!\!\!\!,\label{eq:GLMB_apprx_pred}
\end{align}
}where $I\subseteq\mathbb{L}$, $\xi\in\Xi$, $I_{+}\subseteq\mathbb{L}_{+}$,
and
\begin{eqnarray}
\!\!\!\!\!\!\!\!\!p_{+}^{(I,\xi,I_{+})}(x,\ell) & \!\!\!\!\!= & \!\!\!\!\!\begin{cases}
p_{B,+}(x,\ell), & \ell\in\mathbb{B}_{+}\\
p_{G,+}^{(I,\xi,I_{+}-\mathbb{B}_{+})}(x,\ell), & \ell\notin\mathbb{B}_{+}
\end{cases},\\
\!\!\!\!\!\!\!\!\!p_{G,+}^{(I,\xi,L)}(x,\ell) & \!\!\!\!\!= & \!\!\!\!\!1_{L}(\ell)\!\left\langle p_{G,+}^{(I,\xi)}\!\left(\{(x,\ell)\}\uplus(\cdot)\right)\!\right\rangle \!(L-\{\ell\}).
\end{eqnarray}
Moreover, if (\ref{eq:GLMB_apprx_pred}) is the prediction density,
then the GLMB filtering density given the multi-object measurement
$Z_{+}$ is \textup{
\begin{align}
\!\!\!\!\!\boldsymbol{\tilde{\pi}}_{+}(\boldsymbol{X}_{+}|Z_{+})\propto\qquad\qquad\qquad\qquad\enskip\qquad\qquad\qquad\qquad\nonumber \\
\Delta(\boldsymbol{X}_{\!+})\!\!\!\sum_{I,\xi,I_{+},\theta_{+}}\!\!\!\tilde{\omega}_{Z+}^{(I,\xi,I_{+},\theta_{+})}\delta_{I_{+\!}}[\mathcal{L}(\boldsymbol{X}_{\!+})]\!\left[\tilde{p}_{Z_{+}}^{(I,\xi,I_{+},\theta_{+})}\right]^{\boldsymbol{X}_{+}}\!\!\!\!,\label{eq:proposal_filtering}
\end{align}
}where $I\subseteq\mathbb{L}$, $\xi\in\Xi$, $I_{+}\subseteq\mathbb{L}_{+},\theta_{+}\in\Theta_{+}$,
and
\begin{eqnarray}
\!\!\!\!\!\!\tilde{\omega}{}_{Z_{+}}^{(I,\xi,I_{+},\theta_{+})} & \!\!\!\!\!= & \!\!\!\!\!\omega_{+}^{(I,\xi,I_{+})}1_{\Theta_{+}(I_{+})}(\theta_{+})\times\nonumber \\
\!\!\!\!\!\! & \!\!\!\!\! & \!\!\!\!\!\left[\eta_{B,Z_{+}}^{(\theta_{+})}\right]^{I_{+}\cap\mathbb{B}_{+}}\left[\tilde{\eta}_{G,Z_{+}}^{(I,\xi,I_{+},\theta_{+})}\right]^{(I_{+}-\mathbb{B}_{+})}\!\!,\label{eq:pred_aprx_weight}\\
\!\!\!\!\!\!\tilde{\eta}_{G,Z_{+}}^{(I,\xi,I_{+},\theta_{+})}(\ell) & \!\!\!\!\!= & \!\!\!\!\!\left\langle p_{G,+}^{(I,\xi,I_{+})}(\cdot,\ell),\varPsi_{Z_{+}}^{(\theta_{+})}(\cdot,\ell)\right\rangle ,\\
\!\!\!\!\!\!\tilde{p}_{Z_{+}}^{(I,\xi,I_{+},\theta_{+})}(x,\ell) & \!\!\!\!\!\propto & \!\!\!\!\!p_{+}^{(I,\xi,I_{+})}(x,\ell)\varPsi_{Z_{+}}^{(\theta_{+})}(x,\ell).
\end{eqnarray}
\end{cor}

\subsubsection*{Update Approximation}

This strategy performs a joint prediction and update, followed by
a GLMB approximation with matching first moment and cardinality distribution
using Proposition 2 of \cite{032}, as summarized in Corollary \ref{cor:upd_apprx}. 
\begin{cor}
\label{cor:upd_apprx}A GLMB that matches the filtering density (\ref{eq:exact_filtering})
in first moment and cardinality distribution is given by \textup{
\begin{align}
\!\!\!\!\widehat{\boldsymbol{\pi}}_{\!+}(\boldsymbol{X}_{+}|Z_{+})\propto\qquad\qquad\qquad\qquad\enskip\qquad\qquad\qquad\qquad\nonumber \\
\Delta(\boldsymbol{X}_{\!+})\!\!\!\!\!\sum_{I,\xi,I_{+},\theta_{+}}\!\!\!\!\omega_{Z_{+}}^{(I,\xi,I_{+},\theta_{+})}\delta_{I_{+\!}}[\mathcal{L}(\boldsymbol{X}_{\!+})]\left[p_{Z_{+}}^{(I,\xi,I_{+},\theta_{+})}\right]^{\boldsymbol{X}_{\!+}}\!\!\!\!,\label{eq:GLMB_apprx_upd}
\end{align}
}where $I\subseteq\mathbb{L}$, $\xi\in\Xi$, $I_{+}\subseteq\mathbb{L}_{+},\theta_{+}\in\Theta_{+}$,
and
\begin{eqnarray}
\!\!\!\!\!\!\!\!\!\!p_{Z_{+}}^{(I,\xi,I_{+},\theta_{+\!})\!}(x,\ell) & \!\!\!\!\!= & \!\!\!\!\!\begin{cases}
p_{B,Z_{+}}^{(\theta_{+})}(x,\ell), & \ell\in\mathbb{B}_{+}\\
p_{G,Z_{+}}^{(I,\xi,I_{+}-\mathbb{B}_{+},\theta_{+})}(x,\ell), & \ell\notin\mathbb{B}_{+}
\end{cases},\\
\!\!\!\!\!\!\!\!\!\!\!p_{G,Z_{+}}^{(I,\xi,L,\theta_{+}\!)\!}(x,\ell) & \!\!\!\!\!= & \!\!\!\!\!1_{\!L}(\ell)\!\left\langle \!p_{G,Z_{+}}^{(I,\xi,\theta_{+\!})\!}\!\left(\{(x,\ell)\}\!\uplus\!(\cdot)\right)\!\right\rangle \!(L\!-\!\{\ell\}).
\end{eqnarray}
\end{cor}
In principle, (\ref{eq:GLMB_apprx_upd}) provides a more accurate
approximation to the multi-object filtering density (\ref{eq:exact_filtering})
than (\ref{eq:proposal_filtering}). However, it is more expensive
to compute due to the joint densities in (\ref{eq:etaG}). Nonetheless,
it is cheaper than propagating (\ref{eq:exact_filtering}), because
the GLMB approximation caps the dimension of the joint densities\footnote{Equations (\ref{eq:comp_stats_01})-(\ref{eq:comp_stats_05}) are
also applicable to the GLMB density.}.

\textit{Remark.} Since the number of components in the multi-object
density grows exponentially over time, truncation of its components
is needed to maintain tractability. Separate prediction and update
implementation is structurally inefficient because it requires two
independent truncations of the multi-object densities. Since the predicted
multi-object density is truncated separately from the update, computations
would be wasted in updating the predicted components that generate
negligible updated components. Algebraically, the prediction and update
can be combined into a single expression to improve efficiency because
truncation of the prediction is no longer needed \cite{007}.

\subsection{Cell Appearance in Mitosis\label{subsec:Cell-appearance}}

\begin{figure}
\begin{centering}
\includegraphics[width=0.45\textwidth]{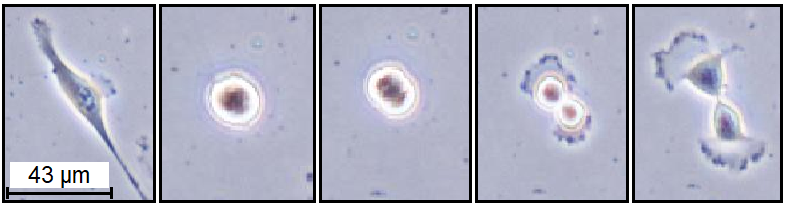}
\par\end{centering}
\caption{A breast cancer cell during mitosis.\label{fig:demo_mitosis}}
\end{figure}

A cell can be in either the normal mode or the mitotic (about to divide)
mode, in which its biological structure changes drastically. This
is manifested via changes the cell's appearance such as shape and
intensity, see Fig. \ref{fig:demo_mitosis}. The proposed multi-object
modeling/estimation framework allows us to exploit the observed cell
appearance to infer the cell modes, which, in turn improves detection
of mitosis events and lineage estimation. 

To capture differences in cell appearance between the two modes, we
augment each cell's unlabeled state vector with a mode variable, i.e.
$x=(\zeta,m)\in\mathbb{X=K}\times\mathbb{M}$, where $\mathbb{K}$
is the kinematic/feature space, and $\mathbb{M=}\{1,2\}$ is the mode
space with `1' and `2' representing the normal and mitotic modes,
respectively. The single object observation $z=(\varrho,\alpha)$
consists of the vector $\varrho$ of kinematic features (e.g. centroid,
velocity), and the vector $\alpha$ of observed appearance features
(e.g. intensity, shapes, or features extracted via a neural network).
To model the dependence of the cell's appearance on the mode, we propose
an observation likelihood function of the form
\begin{equation}
g(\varrho,\alpha|\zeta,m,\ell)=g^{(k)}(\varrho|\zeta,\ell)g^{(a)}(\alpha|\zeta,m,\ell),\label{eq:Gauss-likelihood-1}
\end{equation}
where the \textit{kinematic likelihood} $g^{(k)}(\varrho|\zeta,\ell)$
is the probability density of the observed kinematic vector, and the
\textit{appearance likelihood} $g^{(a)}(\alpha|\zeta,m,\ell)$ is
the probability density of the observed appearance vector. Note that
$g^{(k)}(\varrho|\zeta,\ell)$ is independent of the mode, while $g^{(a)}(\alpha|\zeta,m,\ell)$
is parameterized by the mode $m$. The relationship between an object's
observed appearance and its state is complex in general, and the appearance
likelihood is usually constructed from training data. 

The time evolution of the mode-augmented state is modeled as a jump-Markov
system \cite{043}. Specifically, our model assumes that if a cell
is currently in survival mode, then at the next time step it will
follow the standard motion model, but could assume either mode. If
it is currently in mitotic mode, then it will divide at the next time
step into daughter cells (and cease to exist). Hence, the kinematic
of an existing cell can only follow the standard motion model. Effectively,
the kinematics of cells are independent of the mode of their generator,
and that their modes are independent of the kinematics of their generator.
As a result, the density $\Phi_{+}^{(c)}$ in the cell division model
(\ref{eq:Transition-with-CD}) has the form
\begin{multline}
\!\!\!\!\Phi_{+}^{(c)\!}(\{(\zeta_{+}^{(1)}\!,m_{+}^{(1)}\!,\ell_{+}^{(1)}),...,(\zeta_{+}^{(c)}\!,m_{+}^{(c)}\!,\ell_{+}^{(c)})\}|(\zeta,m,\ell))=\\
\!\!\!\varphi_{+}^{(c)}(\{(\zeta_{+}^{(1)}\!,\ell_{+}^{(1)}),...,(\zeta_{+}^{(c)}\!,\ell_{+}^{(c)})\}|\zeta,\ell)\prod_{i=1}^{c}\vartheta^{(c)}(m_{+}^{(i)}|m,\ell),\!\!\!
\end{multline}
where $\varphi_{+}^{(c)}\left(\cdot\mid\zeta,\ell\right)$ is the
joint density of kinematics and labels of the $c$ cells generated
at the next time, and $\vartheta^{(c)}(m_{+}^{(i)}|m,\ell)$ is the
probability that the generated cell with label $\ell_{+}^{(i)}$ takes
on mode $m_{+}^{(i)}$. \textcolor{black}{Note that for} $c=1,$ $\vartheta^{(1)}(m_{+}^{(1)}|m,\ell)$
is the mode transition probability of the cell with label $\ell$,
if it survives.

\subsection{Extension to Unknown Background and Birth Parameters\label{subsec:UnknownBG}}

To address the unknown clutter rate, we adopt the strategy proposed
in \cite{043}, which treats clutter as an independent type of objects.
In our context, a clutter object cannot divide and only take on one
mode (normal mode) while its kinematic state is uniformly distributed
over the observation region. To address the unknown cell detection
rate, we augment $d\in[0,1]$ to the state of the object, i.e. $\boldsymbol{x}=(\zeta,m,d,\ell)$,
and define $P_{D}(\zeta,m,d,\ell)\triangleq d$ \cite{009}. The clutter
detection rate is assumed to be a constant $P_{D}^{(0)}$ and the
cell detection rate is modeled with a beta distribution which is propagated
as in \cite{009}.

A static LMB birth model can be slow in initiating tracks. This can
be alleviated by an adaptive LMB birth model that uses previous measurements
to construct the LMB parameters at the current time step \cite{030,040}.
Additionally, as new cells tend to enter the tracking region from
the edges, the existence probability of new births can be adjusted
accordingly, i.e. higher toward the edge of the image and low near
the center. 

\section{Cell Tracking Filter Implementations\label{sec:Cell-Tracking-Filter}}

This section details the implementations of the cell tracking filters
discussed in Subsection \ref{subsec:Multi-object-filtering}. The
proposed spawning model allows an object to generate up to $C$ objects,
but for cell tracking we only need the special case $C=2$. Nonetheless,
the solutions presented here readily extend to larger $C$. 

The most pressing implementation issue is the exponential growth in
the number of terms/components of the filtering densities. To maintain
tractability, we truncate the insignificant (low-weight) components,
which minimizes the $L_{1}$-error from the original GLMB \cite{006}\footnote{This also holds for densities of the form (\ref{eq:general_init}),
using the same line of arguments as Proposition 5 of \cite{006},
noting that $\left\langle p^{(\xi)}\right\rangle (L)=1.$}. In Subsection \ref{subsec:Ranked-assignment}, we formulate the
ranked assignment problem for truncating the multi-object densities
in Corollary \ref{cor:pred_apprx} (prediction approximation) and
Corollary \ref{cor:upd_apprx}/Proposition \ref{prop:exact_update}
(update approximation/exact filtering). Since the problem size is
very large, traditional ranked assignment solutions \cite{006} are
not tractable while the Gibbs sampler of \cite{007} is not directly
applicable due to the conditional parent-daughter dependence. In Subsection
\ref{subsec:Block-Gibbs-sampling}, we propose a block Gibbs sampling
solution that can accommodate this dependence. Computing the single-object
densities of the resulting GLMB is discussed in Subsection \ref{subsec:single_object_models}.

\subsection{Ranked Assignment Problem\label{subsec:Ranked-assignment}}

\begin{figure*}
\begin{align}
\!\!\!\!\lambda_{i}^{(I,\xi)}(j) & \triangleq\begin{cases}
\eta_{G,+}^{(\ell_{i},\xi)}(\emptyset), & \!\ell_{i}\in I;j\in\mathbb{N}_{+};j^{(3)}=-1,\\
\eta_{G,+}^{(\ell_{i},\xi)}(\{\ell_{i}\})\eta_{G,Z_{+}}^{(\ell_{i},\xi,\{\ell_{i}\})}(j^{(3)},\ell_{i}), & \!\ell_{i}\in I;j\in\mathbb{N}_{+};j^{(3)}>-1,\\
\eta_{G,+}^{(\ell_{i},\xi)}(\{\ell_{i}^{(1)},\ell_{i}^{(2)}\})\prod_{q=1}^{2}\eta_{G,Z_{+}}^{(\ell_{i},\xi,\{\ell_{i}^{(1)},\ell_{i}^{(2)}\})}(j^{(q)},\ell_{i}^{(q)}), & \!\ell_{i}\in I;j\in\mathbb{D}_{+};\left(\left(j^{(1)}\neq j^{(2)}\right)\textrm{or}\left(j^{(1)}=j^{(2)}=0\right)\right),\\
1-r_{B,+}(\ell_{i}), & \!\ell_{i}\in\mathbb{B_{+}};j\in\mathbb{N}_{+};j^{(3)}=-1,\\
r_{B,+}(\ell_{i})\langle p_{+}^{(B)}(\cdot,\ell_{i})\psi_{Z_{+}}^{(j^{(3)})}(\cdot,\ell_{i})\rangle, & \!\ell_{i}\in\mathbb{B}_{+};j\in\mathbb{N}_{+};j^{(3)}>-1,\\
0 & \!\textrm{otherwise}.
\end{cases}\label{eq:eta-Gibbs}
\end{align}
\end{figure*}
Since the maximum number of daughter cells is 2 ($C=2$), we denote
the set of possible labels generated at the next time from any $\ell\in\mathbb{L}$
as $\{\ell^{(1)},\ell^{(2)},\ell^{(3)}\}$, where $\ell^{(1)}=(\ell,k+1,2,1)$,
$\ell^{(2)}=(\ell,k+1,2,2)$, are the daughter labels, and $\ell^{(3)}=\ell$
is the parent label (see Subsection \ref{subsec:Multi-object-models}). 

Note from the recursions (\ref{eq:proposal_filtering}) and (\ref{eq:GLMB_apprx_upd})
that each GLMB component (indexed by) $(I,\xi)$ generates, at the
next time, a set of \textquotedblleft children\textquotedblright{}
components $(I,\xi,I_{+},\theta_{+})$. For a prior component $(I,\xi)$,
let us enumerate $Z_{+}=\{z_{1:M}\}$, $I=\{\ell_{1:R}\}$, $\mathbb{B}_{+}=\{\ell_{R+1:P}\}$,
and represent each pair $(I_{+},\theta_{+})\in\mathcal{F}(\mathbb{L}_{+})\times\Theta_{+}$
by $P\times3$ matrix $\gamma$, called an \textit{extended association
map}, defined as
\begin{equation}
\gamma_{i,q}=\begin{cases}
\theta_{+}(\ell_{i}^{(q)}), & \ell_{i}^{(q)}\in I_{+}\\
-1, & \textrm{otherwise }
\end{cases}.\label{eq:gamma_theta_I}
\end{equation}
We use the notation $\gamma_{i}$ for the $i$-th row of $\gamma$.
In this representation $\gamma_{i,q}=-1$ means $\ell_{i}^{(q)}$
does not exist, $\gamma_{i,q}=0$ means $\ell_{i}^{(q)}$ exists but
not detected, and $\gamma_{i,q}>0$ means $\ell_{i}^{(q)}$ exists
and generates the measurement indexed by $\gamma_{i,q}$. Since a
parent cell cannot co-exist with its daughters, each $\gamma_{i}\in\mathbb{D}_{+}\uplus\mathbb{N}_{+}$,
and $\gamma_{i}\in\mathbb{N}_{+}$ for $i\in\{R+1\colon P\}$, where
\[
\mathbb{D}_{+}=\{0\colon M\}^{2}\times\{-1\},
\]
(i.e. division occurs, daughters exist but not the parent) and
\[
\mathbb{N}_{+}=\{-1\}\times\{-1\}\times\{-1\colon M\},
\]
(i.e. no division). $\gamma$ also inherits the positive 1-1 property,
i.e. there are no distinct $(i,q),(i',q')$ with $\gamma_{i,q}=\gamma_{i',q'}>0$. 

Let $\Gamma$ denote the set of all extended association maps, i.e.
$P\times3$ matrices that are positive 1-1 with $\gamma_{i}\in\mathbb{D}_{+}\uplus\mathbb{N}_{+}$,
$i\in\{1\colon R\}$ and $\gamma_{i}\in\mathbb{N}_{+}$, $i\in\{R+1\colon P\}$.
Then, for any $\gamma\in\Gamma$, we recover $(I_{+},\theta_{+})$
by
\[
I_{+}=\{\ell_{i}^{(q)}\in\mathbb{G}_{+}(I)\uplus\mathbb{B}_{+}:\gamma_{i,q}\geq0\},\ \ \theta_{+}(\ell_{i}^{(q)})=\gamma_{i,q}.
\]
Hence, there is a 1-1 correspondence between $\Theta_{+}(I_{+})$
and $\Gamma$, with $1_{\Gamma}(\gamma)=1_{\Theta_{+}(I_{+})}(\theta_{+})$.
Consequently, selecting the significant children of component $(I,\xi)$
amounts to selecting extended association maps with significant weights
as per (\ref{eq:pred_aprx_weight}) for the prediction approximation,
or (\ref{eq:exact_upd_weight}) for the update approximation. 

\subsubsection*{Prediction Approximation}

Since we are using a GLMB approximation, it can be shown that the
weight (\ref{eq:pred_aprx_weight}) takes the form (for completeness
details are given in Appendix \ref{subsec:Equivalence-of-apprx-pred-weight})
\begin{multline}
\!\!\!\!\tilde{\omega}{}_{Z_{+}}^{(I,\xi,I_{+},\theta_{+})}\propto1_{\Theta_{+}(I_{+})}(\theta_{+})[\eta_{G,Z_{+}}^{(\cdot,\xi,\theta_{+},I_{+})}]^{I}[1-r_{B,+}(\cdot)]{}^{\mathbb{B}_{+}-I_{+}}\\
\begin{array}{c}
\times\left[r_{B,+}(\cdot)\!\int p_{+}^{(B)}(x,\cdot)\psi_{Z_{+}}^{(\theta_{+}(\cdot))}(x,\cdot)dx\right]^{\!I_{+}\cap\mathbb{B}_{+}},\end{array}\label{eq:pred_aprx_weight_equivalence}
\end{multline}
where
\begin{eqnarray}
\!\!\!\!\!\!\!\!\!\!\eta_{G,Z_{+}}^{(\ell,\xi,\theta_{+},I_{+})}\!\!\!\! & = & \!\!\!\!\eta_{G,+}^{(\ell,\xi)}(I_{+}\cap\mathbb{G}_{+}(\ell))\times\\
\!\!\!\!\!\!\!\!\!\!\!\! &  & \!\!\!\!\left[\eta_{G,Z_{+}}^{(\ell,\xi,I_{+}\cap\mathbb{G}_{+}(\ell))}(\theta_{+}(\cdot),\cdot)\right]^{I_{+}\cap\mathbb{G}_{+}(\ell)},\nonumber \\
\!\!\!\!\!\!\!\!\!\!\eta_{G,+}^{(\ell,\xi)}(L)\!\!\!\! & = & \!\!\!\!\langle q_{G,+}^{(\ell,\xi)}\rangle(L),\\
\!\!\!\!\!\!\!\!\!\!\eta_{G,Z_{+}}^{(\ell,\xi,L)}(j,u)\!\!\!\! & = & \!\!\!\!\langle p_{G,+}^{(\ell,\xi,L)}(\cdot,u),\psi_{Z_{+}}^{(j)}(\cdot,u)\rangle,\\
\!\!\!\!\!\!\!\!\!\!p_{G,+}^{(\ell,\xi,L)}(x,u)\!\!\!\! & = & \!\!\!\!1_{L}(u)\!\left\langle p_{G,+}^{(\ell,\xi)}\!\left(\{(x,u)\}\uplus(\cdot)\right)\!\right\rangle \!(L-\{u\}),\\
\!\!\!\!\!\!\!\!\!\!p_{G,+}^{(\ell,\xi)}(\boldsymbol{Y})\!\!\!\! & = & \!\!\!\!q_{G,+}^{(\ell,\xi)}(\boldsymbol{Y})/\eta_{G,+}^{(\ell,\xi)}(\mathcal{L}(\boldsymbol{Y})),\\
\!\!\!\!\!\!\!\!\!\!q_{G,+}^{(\ell,\xi)}(\boldsymbol{Y})\!\!\!\! & = & \!\!\!\!\int\!p^{(\xi)\!}(x,\ell)\boldsymbol{f}_{+}^{(G)\!}(\boldsymbol{Y}\!\cap\mathbb{X\!}\times\!\mathbb{G}_{+}(\ell)|x,\ell)dx.
\end{eqnarray}
Further, using %
\begin{comment}
the relationship between $(I_{+},\theta_{+})$ and $\gamma$ in
\end{comment}
{} (\ref{eq:gamma_theta_I}), we can write the weight $\tilde{\omega}_{Z_{+}}^{(I,\xi,I_{+},\theta_{+})}$
in (\ref{eq:pred_aprx_weight_equivalence}) as a function of $\gamma$.
\begin{prop}
\label{prop:pred-aprx-weight-lambda}For each $i\in\{1\colon P\}$
and triplet $j\in\mathbb{D}_{+}\uplus\mathbb{N}_{+}$, define $\lambda_{i}^{(I,\xi)}(j)$
by (\ref{eq:eta-Gibbs}). Then for any $(I_{+},\theta_{+})\in\mathcal{F}(\mathbb{L}_{+})\times\Theta_{+}$
and its equivalent representation $\gamma\in\Gamma$,
\begin{equation}
\tilde{\omega}_{Z_{+}}^{(I,\xi,I_{+},\theta_{+})}=\omega^{(I,\xi)}1_{\Gamma}(\gamma)\prod_{i=1}^{P}\lambda_{i}^{(I,\xi)}(\gamma_{i}).\label{eq:block-gibbs-w-vs-eta}
\end{equation}
\end{prop}
\begin{comment}
$\begin{alignedat}{1}\eta_{i}^{(I,\xi)}(j)\triangleq\begin{cases}
\eta_{G,+}^{(\ell_{i},\xi)}(\emptyset), & \ell_{i}\in I;j=(-1,-1,-1)\\
\eta_{G,+}^{(\ell_{i},\xi)}(\{\ell_{i}\})\eta_{G,Z_{+}}^{(\ell_{i},\xi,\{\ell_{i}\})}(j^{(3)},\ell_{i}), & \ell_{i}\in I;j^{(3)}\geq0\\
\eta_{G,+}^{(\ell_{i},\xi)}(\{\ell_{i}^{(1)},\ell_{i}^{(2)}\})\prod_{q=1}^{2}\eta_{G,Z_{+}}^{(\ell_{i},\xi,\{\ell_{i}^{(1)},\ell_{i}^{(2)}\})}(j^{(q)},\ell_{i}^{(q)}), & \ell_{i}\in I;j\in\mathbb{M}_{+};\left(\left(j^{(1)}\neq j^{(2)}\right)or\left(j^{(1)}=j^{(2)}=0\right)\right)\\
1-r_{B,+}(\ell_{i}), & \ell_{i}\in\mathbb{B_{+}};j=(-1,-1,-1)\\
r_{B,+}(\ell_{i})\langle p_{+}^{(B)}(\cdot,\ell_{i})\psi_{Z_{+}}^{(j^{(3)})}(\cdot,\ell_{i})\rangle, & \ell_{i}\in\mathbb{B}_{+};j^{(3)}\geq0.\\
0 & \textrm{otherwise}.
\end{cases}\end{alignedat}
$
\end{comment}

Hence, for a given component $\left(I,\xi\right)$ the problem of
selecting its $T$ children with highest weights according to (\ref{eq:block-gibbs-w-vs-eta})
is a ranked assignment problem with cost $C_{i,j}=\lambda_{i}^{(I,\xi)}(j)$,
$i\in\{1\colon P\}$, $j\in\mathbb{D}_{+}\uplus\mathbb{N}_{+}$. Note
that for each $i$, there are $(M+1)^{2}+M+2$ candidate $j$'s. %
\begin{comment}
Note that for each $i\in\{1\colon R\}$, there are $(M+1)^{2}+M+2$
candidate $j$'s, and for each $i\in\{R+1\colon P\}$ there are only
$M+2$ candidate $j$'s
\end{comment}
This ranked assignment problem can be solved using Murty's algorithm
and variants with complexity $\mathcal{O}(T(2P+M^{2}){}^{3})$ \cite{027,027a,027b},
which is still very prohibitive even for a moderate number of cells.
A cheaper alternative is to sample from (\ref{eq:block-gibbs-w-vs-eta})
as detailed in Subsection \ref{subsec:Block-Gibbs-sampling}. 

\subsubsection*{Update Approximation/Exact Filtering Density}

A similar ranked assignment problem can be formulated for truncating
the multi-object filtering density (\ref{eq:exact_filtering}) and
its GLMB approximation, by expressing the weight $\omega_{Z_{+}}^{(I,\xi,I_{+},\theta_{+})}$
(\ref{eq:exact_upd_weight}) as a function of the extended association
map. However, computing the cost $C_{i,j}$ , $i\in\{1\colon P\}$,
$j\in\mathbb{D}_{+}\uplus\mathbb{N}_{+}$ for this ranked assignment
problem is expensive because it involves operating on the joint densities
of the cells. Hence, solving the resulting ranked assignment problem
is impractical when the number of measurements $M$ is large. To circumvent
this computational problem, we propose to sample the extended association
maps from (\ref{eq:block-gibbs-w-vs-eta}) to generate the significant
children components $(I,\xi,I_{+},\theta_{+})$, and recompute their
weights via (\ref{eq:exact_upd_weight}). The rationale is that $[\tilde{\eta}_{G,Z_{+}}^{(I,\xi,I_{+},\theta_{+})}]^{(I_{+}-\mathbb{B}_{+})}$
in (\ref{eq:pred_aprx_weight}) was designed to approximate $\eta_{G,Z_{+}}^{(I,\xi,\theta_{+})}(I_{+}-\mathbb{B}_{+})$
in (\ref{eq:exact_upd_weight}). Hence, if $(I,\xi,I_{+},\theta_{+})$
has a significant $\tilde{\omega}_{Z_{+}}^{(I,\xi,I_{+},\theta_{+})}$
it also has a significant $\omega_{Z_{+}}^{(I,\xi,I_{+},\theta_{+})}$,
even though these values may differ.%

\subsection{Block Gibbs Sampling\label{subsec:Block-Gibbs-sampling}}

This section presents a technique for sampling extended association
maps from the discrete probability distribution $\pi$ given by
\begin{equation}
\pi(\gamma)\propto1_{\Gamma}(\gamma)\prod_{i=1}^{P}\lambda_{i}^{(I,\xi)}(\gamma_{i}).\label{eq:Gibbs-proposal}
\end{equation}
In particular, we use a block Gibbs sampler to generate $\gamma$
row by row, via a Markov chain with transition kernel 
\begin{equation}
\pi(\gamma'|\gamma)=\prod_{n=1}^{P}\pi_{n}(\gamma_{n}^{\prime}|\gamma_{1:n-1}^{\prime},\gamma_{n+1:P}),
\end{equation}
where each conditional $\pi_{n}\left(\cdot|\cdot\right)$\textit{
}is given by 
\begin{equation}
\pi_{n}\left(\gamma_{n}^{\prime}|\gamma_{1:n-1}^{\prime},\gamma_{n+1:P}\right)=\frac{\pi\left(\gamma_{1:n}^{\prime},\gamma_{n+1:P}\right)}{\sum_{\gamma_{n}}\!\pi\left(\gamma_{1:n-1}^{\prime},\gamma_{n},\gamma_{n+1:P}\right)}.
\end{equation}
The following proposition provides closed form expressions for the
conditionals that can be computed/sampled at low cost. The proof follows
that of Proposition 3 in \cite{007} and is provided in Appendix \ref{sec:Proof-of-Proposition-6}
for completeness. %
\begin{comment}
\begin{lem}
Let $\bar{n}\triangleq\{1:P\}-\{n\}$, $\gamma_{\bar{n}}=(\gamma_{1:n-1},\gamma_{n+1:P})$
and $\Gamma(\bar{n})$ be the set of valid $\gamma_{\bar{n}}$ with
positive 1-1 property (no any distinct pairs $(i,j)$ where $i,j\in\bar{n}$
such that $\gamma_{i}^{(q)}=\gamma_{j}^{(r)}>0$, with $q,r\in\{1,2,3\}$.
Then for any valid $\gamma$ we have
\begin{eqnarray}
1_{\Gamma}(\gamma) & = & 1_{\Gamma(\bar{n})}(\gamma_{\bar{n}})\prod_{i\in\bar{n}}\prod_{(q,r)=(1,1)}^{(3,3)}(1-1_{\{1:M\}}(\gamma_{n}^{(q)})\delta_{\gamma_{n}^{(q)}}[\gamma_{i}^{(r)}])\nonumber \\
 & = & 1_{\Gamma(\bar{n})}(\gamma_{\bar{n}})\prod_{i\in\bar{n}}\Upsilon_{\{1:M\}}(\gamma_{n},\gamma_{i}).
\end{eqnarray}
\end{lem}
\begin{prop}
$\pi_{n}(\gamma_{n}|\gamma_{\bar{n}})\propto\begin{cases}
\eta_{n}^{(I,\xi)}(\gamma_{n}) & \textrm{if }\gamma\in\Gamma,\gamma_{n,q}\neq\gamma_{i,r},\forall\gamma_{n,q}>0,i\in\bar{n},q,r\in\{1,2,3\}\\
0 & \textrm{otherwise}
\end{cases}.$
\end{prop}
\end{comment}

\begin{prop}
\label{prop:conditional-sampling-density}Given $n\in\{1\colon P\}$
and $\gamma_{\bar{n}}=(\gamma_{1:n-1},\gamma_{n+1:P})$,
\begin{equation}
\!\pi_{n}(\gamma_{n}|\gamma_{\bar{n}})\propto\begin{cases}
0, & \!\!\!\!\!\begin{array}{l}
\textrm{if any positive entry of \ensuremath{\gamma_{n}} }\\
\textrm{coincides with those of }\gamma_{\bar{n}}
\end{array}\\
\lambda_{n}^{(I,\xi)}(\gamma_{n}), & \!\!\textrm{otherwise}
\end{cases}\!\!.\!\!\label{eq:Gibbs-n-proposal}
\end{equation}
\end{prop}
All iterates of the proposed block Gibbs sampler, summarized in Algorithm
\ref{alg:Block-Gibbs-sampling.}, are extended association maps\textit{.}
Its convergence property is given in the following proposition (the
proof follows that of Proposition 4 in \cite{007} and is provided
in Appendix \ref{sec:Proof-of-Proposition-7}).
\begin{algorithm}
\noindent \textbf{$\quad$Input: $\gamma^{(1)}$},\textbf{ $T$},\textbf{
$\lambda^{(I,\xi)}=\left[\lambda_{i}^{(I,\xi)}(j\in\mathbb{N}_{+}\uplus\mathbb{D}_{+})\right]$}

\noindent \textbf{$\quad$Output: $\gamma^{(1)},...,\gamma^{(T)}$}

\noindent \textbf{\rule[0.5ex]{1\columnwidth}{0.5pt}}

\noindent \textbf{}%
\begin{tabular*}{0.8\columnwidth}{@{\extracolsep{\fill}}l}
\textbf{$\quad$for $t=2:T$}\tabularnewline
\textbf{$\quad$$\quad$for $i=1:P$}\tabularnewline
\textbf{$\quad$$\quad$$\quad$for $j\in\mathbb{N}_{+}\uplus\mathbb{D}_{+}$}$\left(\left|\mathbb{N}_{+}\uplus\mathbb{D}_{+}\right|=(M+1)^{2}+M+2\right)$\tabularnewline
\textbf{$\quad\quad\quad\quad$if }any positive entry of\textbf{ $j$
}in\textbf{ $\left[\gamma_{1:i-1}^{(t)},\gamma_{i+1:P}^{(t-1)}\right]$}\tabularnewline
\textbf{$\quad\quad\quad\quad$$\quad\lambda_{i}^{(I,\xi)}(j):=0$}\tabularnewline
\textbf{$\quad\quad\quad\quad$end}\tabularnewline
\textbf{$\quad\quad\quad$end}\tabularnewline
$\quad\quad\quad\gamma_{i}^{(t)}\sim\textrm{Categorical}(\mathbb{N}_{+}\uplus\mathbb{D}_{+},\lambda_{i}^{(I,\xi)})$\tabularnewline
\textbf{$\quad\quad$end}\tabularnewline
\textbf{$\quad$$\quad\gamma^{(t)}:=[\gamma_{1}^{(t)};...;\gamma_{P}^{(t)}]$}\tabularnewline
\textbf{$\quad$end}\tabularnewline
\end{tabular*}

{\small{}\caption{Block Gibbs sampling.\label{alg:Block-Gibbs-sampling.}}
}{\small\par}
\end{algorithm}
\begin{prop}
\label{prop:Gibbs-convergence}Starting with any $\gamma\in\Gamma$,
the block Gibbs sampler, defined the conditionals in (\ref{eq:Gibbs-n-proposal}),
converges to the stationary distribution (\ref{eq:Gibbs-proposal})
at an exponential rate. Specifically, let $\pi^{j}$ denote the $j^{th}$
power of the transition matrix then
\[
\max_{\lambda,\lambda'\in\Lambda}\left(\left|\pi^{j}(\gamma'|\gamma)-\pi(\gamma')\right|\right)\leq\left(1-2\beta\right)^{\left[\frac{j}{2}\right]},
\]
where $\beta\triangleq\min_{\gamma,\gamma'\in\Gamma}\pi^{2}(\gamma'|\gamma)>0$
is the least likely 2-step transition probability.
\end{prop}
Sampling $\gamma'_{n}\sim\pi_{n}(\cdot|\gamma'_{1:n-1},\gamma'_{n+1:P})$
takes $\mathcal{O}\left(PM^{2}\right)$ operations since the complexity
of categorical sampling is linear in the number of categories. Consequently,
following \cite{007}, the $T$-best solutions are sampled via the
block Gibbs sampler with a complexity of $\mathcal{O}\left(TP^{2}M^{2}\right)$. 

Overall, to sample all components $\left(I,\xi,I_{+},\theta_{+}\right)$
with significant weights according to (\ref{eq:block-gibbs-w-vs-eta}),
we sample $\left(I,\xi\right)$ from $\pi\left(I,\xi\right)\propto\omega^{\left(I,\xi\right)}$,
and then for each $\left(I,\xi\right)$, we sample $\gamma$ (and
hence $\left(I_{+},\theta_{+}\right)$) from (\ref{eq:Gibbs-proposal})
via the described Gibbs sampler.

In addition to the complexity of the block Gibbs sampler, the exact
filter and its update approximation require $T$ operations to compute
each component weight (\ref{eq:exact_upd_weight}). For the exact
filter, the dimension of the joint object state grows when new objects
appear/generated because objects are dependent on each other. Thus,
computing the filtering density involves operating on very high dimensional
spaces. These computations are reduced in the update approximation
strategy since the joint densities are marginalized to form independent
single-object densities at the end of each filtering cycle.

\begin{comment}
\begin{rem}
Due to the nature of permutation, the number of possible assignments
$N$ is super exponential in terms of the number of measurements $M$
and the number of maximum generated cardinality $C$. However, via
applying the measurement gating technique presented in \cite{030}
and the nature of cells division which typically involves $C=2$,
the number of valid permutations reduces significantly which allows
the sampler to perform efficiently.
\end{rem}
\end{comment}

\subsection{State Density Propagation\label{subsec:single_object_models}}

Since we are using a Jump-Markov model, the initial joint object densities
are separable in kinematic and mode, i.e. $p_{0}^{(\xi)}(\{(\zeta_{1},m_{1},\ell_{1}),...,(\zeta_{n},m_{n},\ell_{n})\})$$=p_{K,0}^{(\xi)}(\{(\zeta_{1},\ell_{1}),...,(\zeta_{n},\ell_{n})\})$$\prod_{i=1}^{n}p_{M,0}^{(\xi)}(m_{i},\ell_{i})$.
\negthinspace{}Consequently, all multi-object densities of subsequent densities are
also separable in kinematic and mode as shown in the following.
\begin{cor}
\label{cor:single-object-jump-Markov}Suppose that the joint density
of each component of the current multi-object filtering density (\ref{eq:general_init})
are separable in kinematic and mode. Then the term $p_{G,+}^{(I,\xi)}(\cdot)$
in Proposition 1 takes the form
\begin{multline}
p_{G,+}^{(\{\ell_{1},...,\ell_{n}\},\xi)}(\boldsymbol{Y})=\!\int p_{K,0}^{(\xi)}(\{(\zeta_{1},\ell_{1}),...,(\zeta_{n},\ell_{n})\})\times\\
\prod_{i=1}^{n}\sum_{c=0}^{C}\!\delta_{\mathbb{G}_{+}^{(c)\!}(\ell_{i})}\!\left[\mathcal{L}(\boldsymbol{Y})\cap\mathbb{G}_{+}(\ell_{i})\right]\times\\
\varphi_{+}^{(c)}\!\left(\mathcal{K}(\boldsymbol{Y})\cap(\mathbb{X}\times\mathbb{G}_{+}(\ell_{i}))\!\mid\!\zeta_{i},\ell_{i}\right)d\zeta_{1:n}\times\\
\prod_{i=1}^{n}\sum_{j\in\mathbb{M}}p_{M,0}^{(\xi)}(j,\ell_{i})\!\sum_{c=0}^{C}\!\delta_{\mathbb{G}_{+}^{(c)\!}(\ell_{i})}\!\left[\mathcal{L}(\boldsymbol{Y})\cap\mathbb{G}_{+}(\ell_{i})\right]\\
\times\prod_{(m_{+},\ell_{+})\in\mathcal{M}^{(c)}(\ell_{i},\boldsymbol{Y})}\vartheta^{(c)}(m_{+}\mid j,\ell_{i}).
\end{multline}
Further, if the initial density is a GLMB with separable form $p_{0}^{(\xi)}(x,\ell)=p_{K,0}^{(\xi)}(\zeta,\ell)p_{M,0}^{(\xi)}(m,\ell)$
then,
\begin{multline}
p_{G,+}^{(I,\xi)}(\boldsymbol{Y})=\prod_{\ell\in I}\sum_{c=0}^{C}\!\delta_{\mathbb{G}_{+}^{(c)\!}(\ell)}\!\left[\mathcal{L}(\boldsymbol{Y})\cap\mathbb{G}_{+}(\ell)\right]\times\\
\langle\varphi_{+}^{(c)}\left(\mathcal{K}(\boldsymbol{Y})\cap(\mathbb{X}\times\mathbb{G}_{+}^{(c)\!}(\ell))\right)\mid\cdot,\ell)p_{K,0}^{(\xi)}(\cdot,\ell),1\rangle\times\\
\sum_{j\in\mathbb{M}}\prod_{(m_{+},\ell_{+})\in\mathcal{M}^{(c)}(\ell,\boldsymbol{Y})}\!\vartheta^{(c)}(m_{+}\mid j,\ell)p_{M,0}^{(\xi)}(j,\ell),
\end{multline}
where, for $\boldsymbol{Y}=\biguplus_{i=1}^{n}\{(\zeta_{i},m_{i},\ell_{i})\}$,
$\mathcal{K}(\boldsymbol{Y})=\biguplus_{i=1}^{n}\{(\zeta_{i},\ell_{i})\}$
and $\mathcal{M}^{(c)}(\ell,\boldsymbol{Y})=\biguplus_{i=1}^{n}\{(m_{i},\ell_{i})\}\cap\left(\mathbb{M}\times\mathbb{G}_{+}^{(c)\!}(\ell)\right)$.
\end{cor}
Under linear Gaussian models, $p_{K,0}^{(\xi)}(\{(\cdot,\ell_{1}),...,(\cdot,\ell_{n})\})$
or $p_{K,0}^{(\xi)}(\cdot,\ell)$ can be propagated analytically using
the Kalman recursion. For non-linear non-Gaussian kinematic models,
extended Kalman filter, unscented Kalman filter or particle filter
can be used. 

\section{Experimental Results\label{sec:Experimental-results}}

This section presents three case studies: a small scale scenario with
simulated detections to benchmark the approximate filters against
the optimal filter (Subsection \ref{subsec:Simulation-experiment});
a large scale scenario of more than 100 cells with synthetic image
sequence to benchmark the approximate filters against well-known cell
trackers (Subsection \ref{subsec:Synthetic-cells-migration}); and
a real sequence of breast cancer cells to demonstrate the viability
of our cheapest solution in real applications (Subsection \ref{subsec:Real-breast-cancer}). 

In all three studies, the number of cells varies with time due to
new independent births, mitosis and deaths, and the multi-object filters
use following system model. A cell's kinematic state is its position-velocity
vector $\zeta=[p_{x},p_{y},\dot{p}_{x},\dot{p}_{y}]$, which follows
a mixture of constant velocity model and free diffusion model, with
transition density 
\begin{align*}
\varphi_{+}^{(1)}\left(\zeta_{+},\ell\mid\zeta,\ell\right) & \!\triangleq\!w_{1}\mathcal{N}(\zeta_{+},F_{1}\zeta,Q_{1})+w_{2}\mathcal{N}(\zeta_{+},F_{2}\zeta,Q_{2}),
\end{align*}
where $\mathcal{N}(\cdot,\bar{\zeta},P)$ is a Gaussian distribution
with mean $\bar{\zeta}$ and co-variance $P$, $w_{1},w_{2}$ are
the mixture weights,
\begin{align*}
F_{1}\!=\!\left[\begin{array}{cc}
I_{2} & I_{2}\\
0_{2} & I_{2}
\end{array}\right]\!\!,\enskip & Q_{1}=\sigma_{v}^{2}\left[\begin{array}{cc}
\frac{1}{4}I_{2} & \frac{1}{2}I_{2}\\
\frac{1}{2}I_{2} & I_{2}
\end{array}\right]\!\!,F_{2}=\left[\begin{array}{cc}
I_{2} & 0_{2}\\
0_{2} & 0_{2}
\end{array}\right]\!\!,
\end{align*}
$Q_{2}=\sigma_{s}F_{2}$, $\sigma_{v}=1$ $\textrm{pixel}/\textrm{frame}^{2}$
and $\sigma_{s}=9$ pixels. This model is motivated by the cell model
provided in the supplementary material of \cite{CTP002}, in which,
cells are assumed to randomly switch between free diffusive (FD) and
directed motion (DM). For mitosis, each daughter cell appears approximately
10 pixels away from the parent's last position. The post-mitosis kinematic
transition density is
\[
\varphi_{+}^{(2)}\!\left(\zeta_{+}^{(1:2)},\mathbb{G}_{+}^{(2)\!}(\ell)\mid\zeta,\ell\right)\!\triangleq\sum_{n=1}^{N}\frac{\mathcal{N}(\zeta_{+}^{(1:2)},F^{(2)}\zeta+d_{n},Q^{(2)})}{N},
\]
where $F^{(2)}=I_{2}\otimes F_{2}$, $Q^{(2)}=I_{2}\otimes Q_{2}$,
$d_{n}=[d_{n}^{(0)},-d_{n}^{(0)}]^{T}$ with $d_{n}^{(0)}=10[\cos(\hat{\theta}+\epsilon\times n),\sin(\hat{\theta}+\epsilon\times n),0,0]$,
$\hat{\theta}$ and $\epsilon$ (degrees) are constants. This mitotic
model is based on the observation in typical cell migration datasets
where daughter cells move in opposite direction at mitosis. Multiple
Gaussian components are used to take into account the uncertainty
in splitting direction of the cells.

The mode transition probabilities are time invariant, given by $\vartheta^{(c)}(m_{+}^{(i)}\mid j,\ell)=p_{sp}$
if $m_{+}^{(i)}=2$ and $\vartheta^{(c)}(m_{+}^{(i)}\mid j,\ell)=1-p_{sp}$
if $m_{+}^{(i)}=1$ (with $p_{sp}=0.03$). The cardinality distribution
given a specific mode is given in Tab. \ref{tab:exp1-mode-vs-card}.

\begin{table}
\caption{Cardinality distribution given a mode.\label{tab:exp1-mode-vs-card}}

\centering{}%
\begin{tabular}{|c|c|c|c|}
\hline 
 & {\small{}$c=0$} & {\small{}$c=1$} & {\small{}$c=2$}\tabularnewline
\hline 
{\small{}$m=1$} & {\small{}0.01} & {\small{}0.98} & {\small{}0.01}\tabularnewline
\hline 
{\small{}$m=2$} & {\small{}0.01} & {\small{}0.09} & {\small{}0.9}\tabularnewline
\hline 
\end{tabular}
\end{table}
The kinematic observation is modeled by the Gaussian likelihood $g^{(k)}(\varrho_{+}|\zeta_{+},\ell_{+})=\mathcal{N}(\varrho_{+},H\zeta_{+},R)$,
where $H=\left[\begin{array}{cc}
I_{2} & 0_{2}\end{array}\right]$, $R=\sigma_{\epsilon}^{2}I_{2}$ and $\sigma_{\epsilon}=2$ pixels.
The mode likelihood is described separately in each experiment.

\subsection{Simulated Detection Experiment\label{subsec:Simulation-experiment}}

\begin{figure}
\begin{centering}
\includegraphics[width=0.45\textwidth]{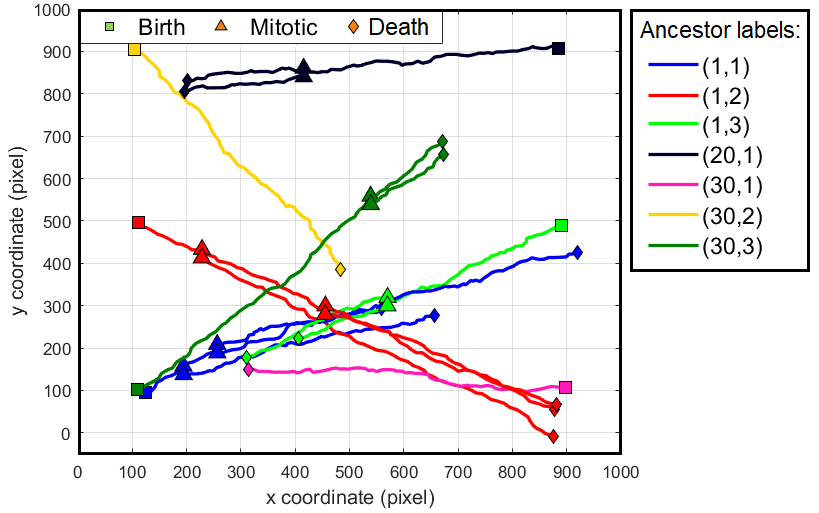}
\par\end{centering}
\caption{True cell trajectories in simulated detection experiment, each distinct
color indicates a family.\label{fig:exp1_truth_traj} }
\end{figure}

In this experiment, cells follow the constant velocity motion, i.e.,
the kinematic model with $w_{1}=1$ and $w_{2}=0$. The mitosis model
has parameters: $N=1$, mean bearing angle of the parent $\hat{\theta}$
and $\epsilon=90^{\circ}$. Ground truth trajectories are shown in
Fig. \ref{fig:exp1_truth_traj}, with a maximum of 12 at any time. 

Each simulated detection is a vector comprising the 2D position and
appearance feature of the cell. These detections are generated with
a detection probability of $0.9$, while clutter is uniformly distributed
with an average rate of $30$. The appearance feature $\alpha=[\alpha_{1},\alpha_{2}]$
is sampled from Beta distributions. Specifically, if this measurement
is generated by: a normal cell then $\alpha_{1}\sim\beta(0.9,0.1)$
and $\alpha_{2}\sim\beta(0.2,0.1)$; a mitotic cell then $\alpha_{1}\sim\beta(0.2,0.1)$
and $\alpha_{2}\sim\beta(0.9,0.1)$. If it is a clutter object then
$\alpha_{1}\sim\beta(0.4,0.1)$ and $\alpha_{2}\sim\beta(0.1,0.1)$.
The mode likelihood is given as $g^{(a)}(\alpha|1)=\alpha_{1}$ and
$g^{(a)}(\alpha|2)=\alpha_{2}$.

\begin{figure}
\begin{centering}
\includegraphics[width=0.45\textwidth]{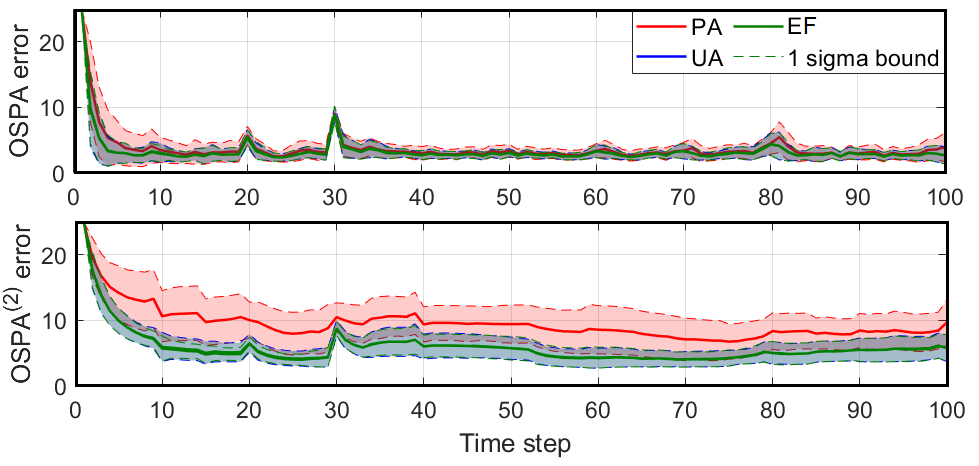}
\par\end{centering}
\caption{Mean OSPA and $\textrm{OSPA}^{\textrm{(2)}}$ errors in simulated
detection experiment. \label{fig:exp1_OSPA} }
\end{figure}
\begin{figure}

\begin{centering}
\includegraphics[width=0.45\textwidth]{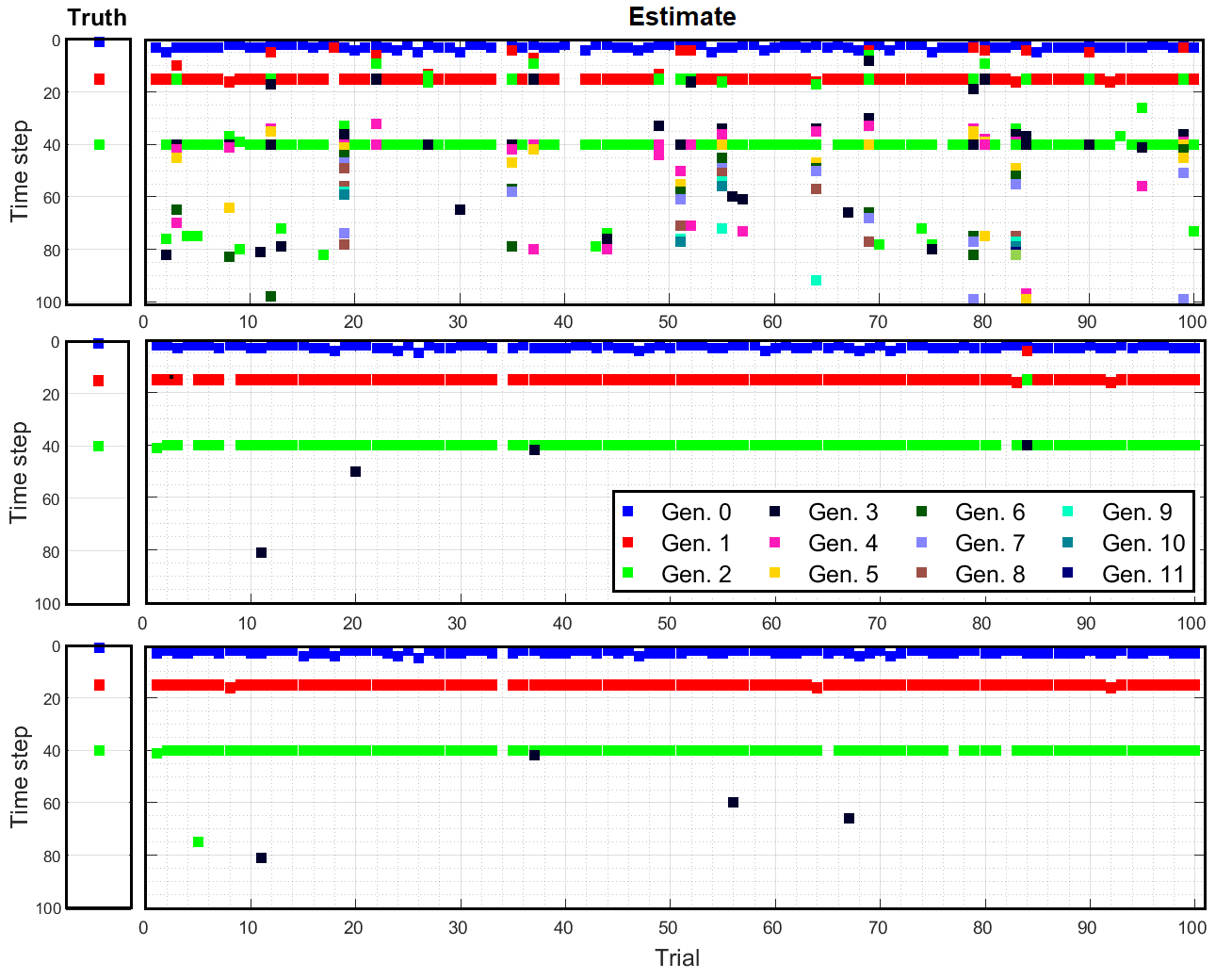}
\par\end{centering}
\caption{Lineage estimation for one cell family (top: PA, middle: UA, bottom:
EF), similar trends are also observed for other families.\label{fig:exp1-lineage} }

\end{figure}

For the purposes of benchmarking the prediction approximation (PA)
and update approximation (UA) against the very expensive implementation
of the exact filter (EF), we assume the clutter rate and detection
probability are known (whereas these are unknown to the trackers in
the next two experiments). All filtering strategies are performed
with a component weight threshold of $10^{-5},$a requested number
of 30000 solutions from the Gibbs sampler, and a maximum number of
30000 components retained. The mean OSPA and $\textrm{OSPA}^{\textrm{(2)}}$
\cite{040} errors over 100 Monte Carlo (MC) trials shown in Fig.
\ref{fig:exp1_OSPA}. The norm-order and cut-off of the OSPA metrics
are set to 1 and 25, respectively (as in \cite{CTP002}), and the
window length for $\textrm{OSPA}^{\textrm{(2)}}$ is set to 20 time
steps. The OSPA errors for the PA, UA and EF are similar due to the
fact that the OSPA does not capture labeling errors. This is confirmed
by the $\textrm{OSPA}^{\textrm{(2)}}$ which shows that the UA and
EF incur a much lower tracking error than the PA, and can be attributed
to more accurate estimation of mitotic events as shown in Fig. \ref{fig:exp1-lineage}.
The error curves for UA and EF are almost identical which shows that
UA is a good approximation of the exact solution.

\begin{figure}
\begin{centering}
\includegraphics[width=0.45\textwidth]{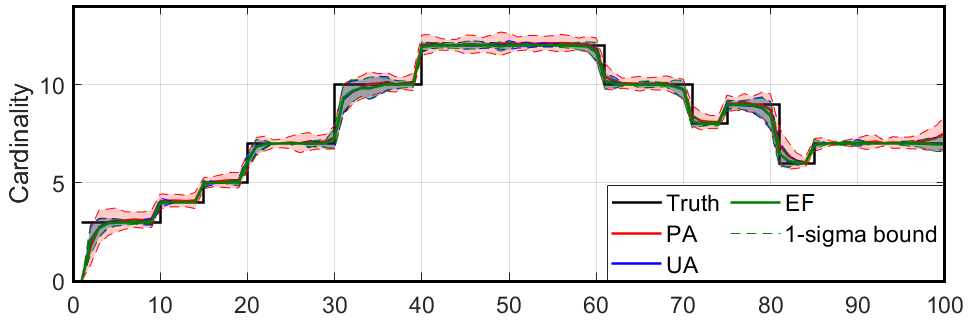}
\par\end{centering}
\begin{centering}
\includegraphics[width=0.45\textwidth]{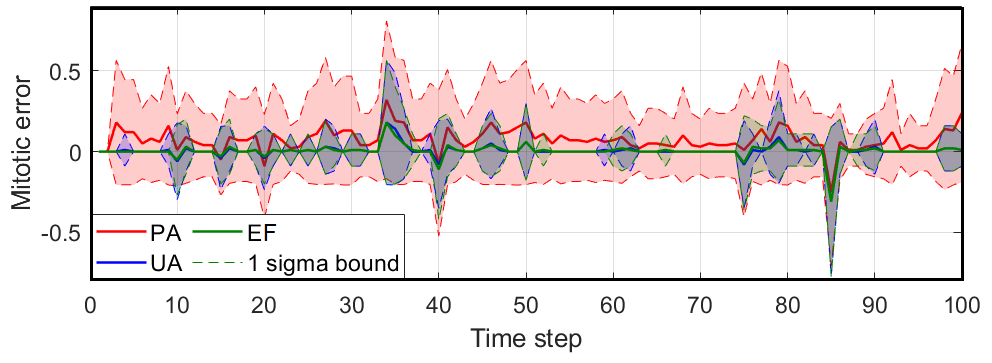}
\par\end{centering}
\caption{Mean estimated cell cardinality (top) and error in mitotic event counts
(bottom) (positive: overestimation, negative: underestimation) in
simulated detection experiment. \label{fig:exp1_card}}
\end{figure}

It can be seen from Fig. \ref{fig:exp1_card} that the PA, UA and
EF correctly estimate the number of cells, but that the UA and EF
exhibit a much lower estimation uncertainty. It further shows that
PA slightly overestimates the number of mitotic events. Due to the
relatively small number of cells, the PA and UA have approximately
similar run times while EF takes significantly longer (more than three
times).


\subsection{Synthetic Cells Migration Sequences\label{subsec:Synthetic-cells-migration}}

\begin{figure}
\begin{centering}
\includegraphics[width=0.45\textwidth]{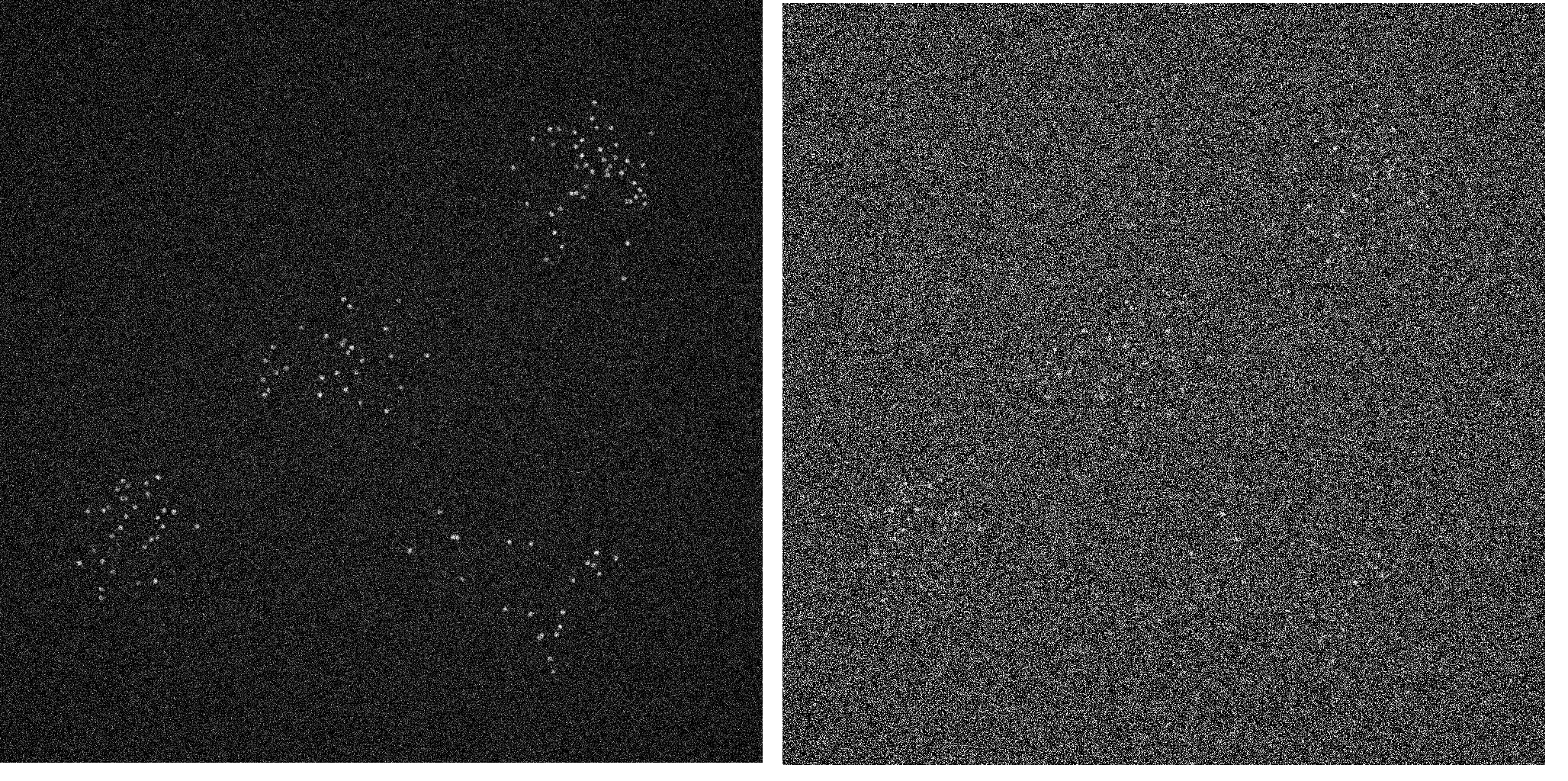}
\par\end{centering}
\caption{Snapshots of the synthetic migration sequence at time step 100 (left:
scenario 1, right: scenario 5).\label{fig:exp2-sample-synthetic} }
\end{figure}

In this experiment, cells initially appear randomly with cardinality
sampled from a Poisson distribution and locations sampled uniformly
within the image. The parameters used to generate the true cell trajectories
are given in Tab. \ref{tab:exp2-truth-gen-param-synthetic}. Parameters
for the kinematic model are $w_{1}=0.3$ and $w_{2}=0.7$. Parameters
for the mitosis model are $N=9$, $\hat{\theta}=0^{\circ}$, and $\epsilon=20^{\circ}$.
Instead of detection sequences, the method in \cite{CTP024} is used
to generate 5 different scenarios, containing fluorescent image sequences
of cell nucleii, and each with a different level of Charge-Coupled
Device noise. Mitotic cells appear with maximum intensity and with
a highly eccentric appearance in mimicking a common characteristic
in the cell division process. Snapshots for scenarios 1 and 5 are
given in Fig. \ref{fig:exp2-sample-synthetic}. From each of the generated
image sequences, the detector proposed in \cite{D014} is used to
extract cell centroids. Based on a distance limit of 5 pixels around
the ground truths, the actual true and false positive rates are tabulated
in Tab. \ref{tab:exp2-pD-clutter}. The intensity of each detected
spot is used as the observed feature of the cell. The likelihood for
the intensity is mode dependent, and $g^{(a)}(\cdot|1)$ is designed
such that its output is high when the intensity is low, while $g^{(a)}(\cdot|2)$,
is high when the intensity is high.

\begin{table}
\caption{Parameters for true cell trajectories generation in synthetic migration
experiment. \label{tab:exp2-truth-gen-param-synthetic}}

\centering{}%
\begin{tabular}{|c|c|}
\hline 
\textbf{\small{}Parameters} & \textbf{\small{}Values}\tabularnewline
\hline 
{\small{}Initial number of cells} & {\small{}20}\tabularnewline
\hline 
{\small{}Sequence length} & {\small{}100}\tabularnewline
\hline 
{\small{}Image size} & {\small{}$1000\times1000$}\tabularnewline
\hline 
{\small{}Poisson rate of birth events} & {\small{}0.1}\tabularnewline
\hline 
{\small{}Probability of death events} & {\small{}0.01}\tabularnewline
\hline 
{\small{}Probability of mitotic events} & {\small{}0.05}\tabularnewline
\hline 
{\small{}DM/FD switching probability} & {\small{}0.3/0.7}\tabularnewline
\hline 
{\small{}Uncertainty of free diffusive motion} & {\small{}10 pixels}\tabularnewline
\hline 
\end{tabular}
\end{table}
\begin{table}
\caption{True/False positive rates of the detection in synthetic experiment.\label{tab:exp2-pD-clutter}}

\centering{}%
\begin{tabular}{|c|c|c|c|c|c|}
\hline 
\multicolumn{1}{|c|}{} & \textbf{\footnotesize{}Scen. 1} & \textbf{\footnotesize{}Scen. 2} & \textbf{\footnotesize{}Scen. 3} & \textbf{\footnotesize{}Scen. 4} & \textbf{\footnotesize{}Scen. 5}\tabularnewline
\hline 
\textbf{\footnotesize{}True Pos. Rate} & {\footnotesize{}0.8473} & {\footnotesize{}0.7752} & {\footnotesize{}0.5727} & {\footnotesize{}0.4566} & {\footnotesize{}0.3862}\tabularnewline
\hline 
\textbf{\textcolor{black}{\footnotesize{}False Pos. Rate}} & {\footnotesize{}0} & {\footnotesize{}0.97} & {\footnotesize{}20.6} & {\footnotesize{}62.78} & {\footnotesize{}105.72}\tabularnewline
\hline 
\end{tabular}
\end{table}
The same filter settings as the previous experiment are used. However,
our filters assume no knowledge of clutter rate and detection parameters.
The unknown detection probability is modeled as a Beta distribution
as per \cite{009}. For unknown clutter rate estimation \cite{043},
the clutter birth rate is set to 0.5 while the clutter surviving and
detection rates are set to 0.9 for all scenarios. Due to the large
number of cells, the EF becomes intractable as it is intensive in
both memory and computations.

\begin{figure}
\begin{centering}
\includegraphics[width=0.45\textwidth]{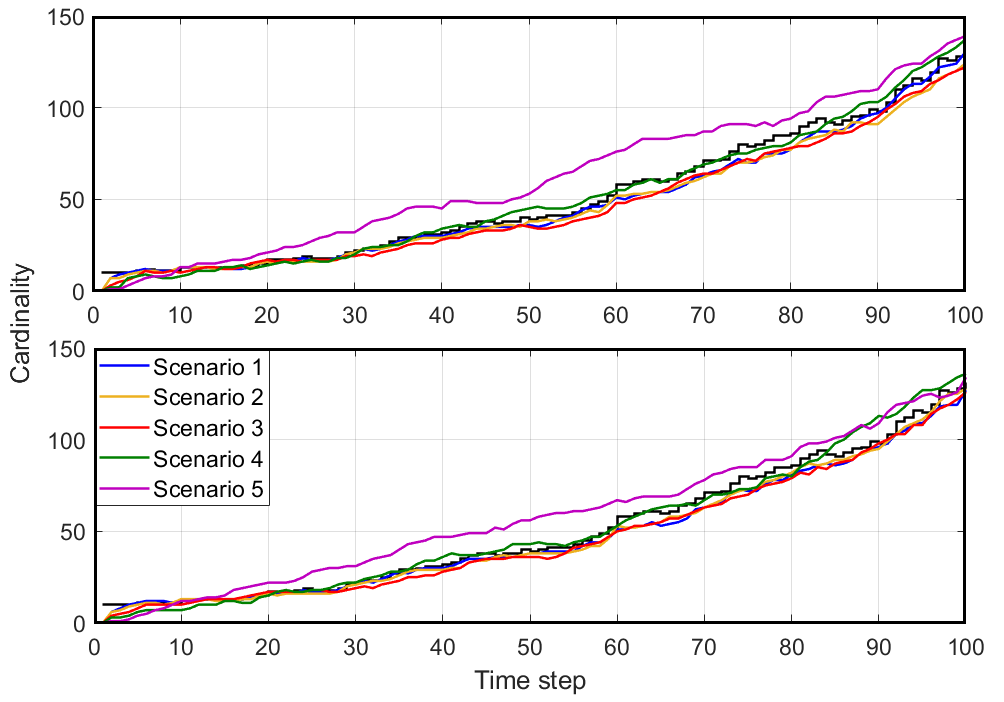}
\par\end{centering}
\caption{Cell cardinality in synthetic migration experiment (top: PA, bottom:
UA).\label{fig:exp2-est-card-synthetic}}
\end{figure}

\begin{figure}
\begin{centering}
\includegraphics[width=0.45\textwidth]{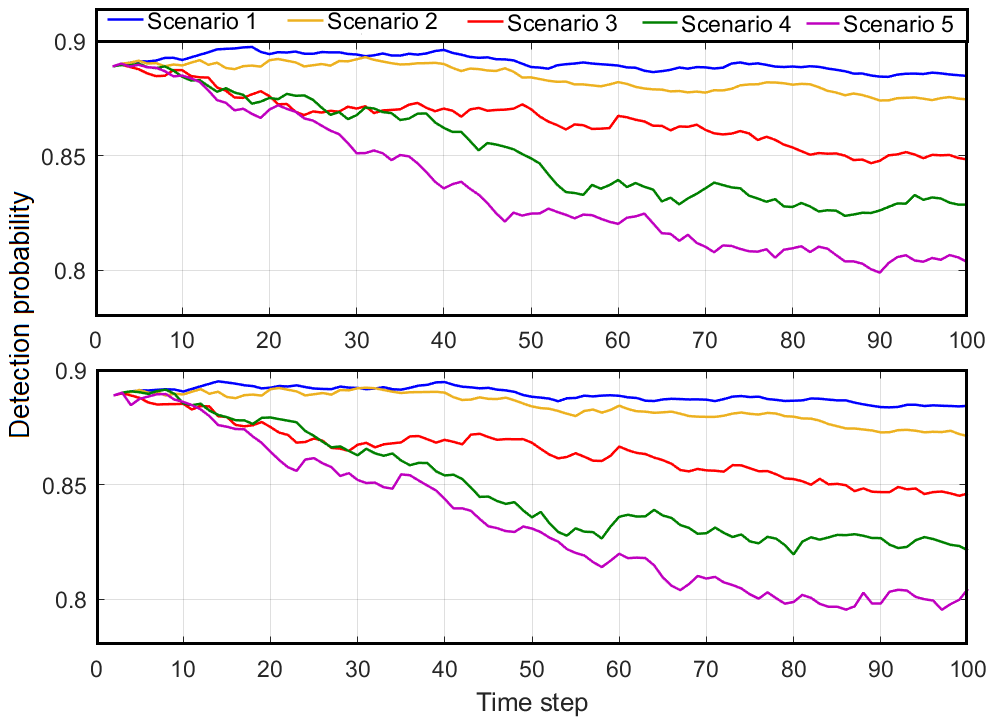}
\par\end{centering}
\caption{Average estimated cell detection probability in synthetic migration
experiment (top: PA, bottom: UA).\label{fig:exp2-pd-synthetic}}
\end{figure}

\begin{figure}
\begin{centering}
\includegraphics[width=0.45\textwidth]{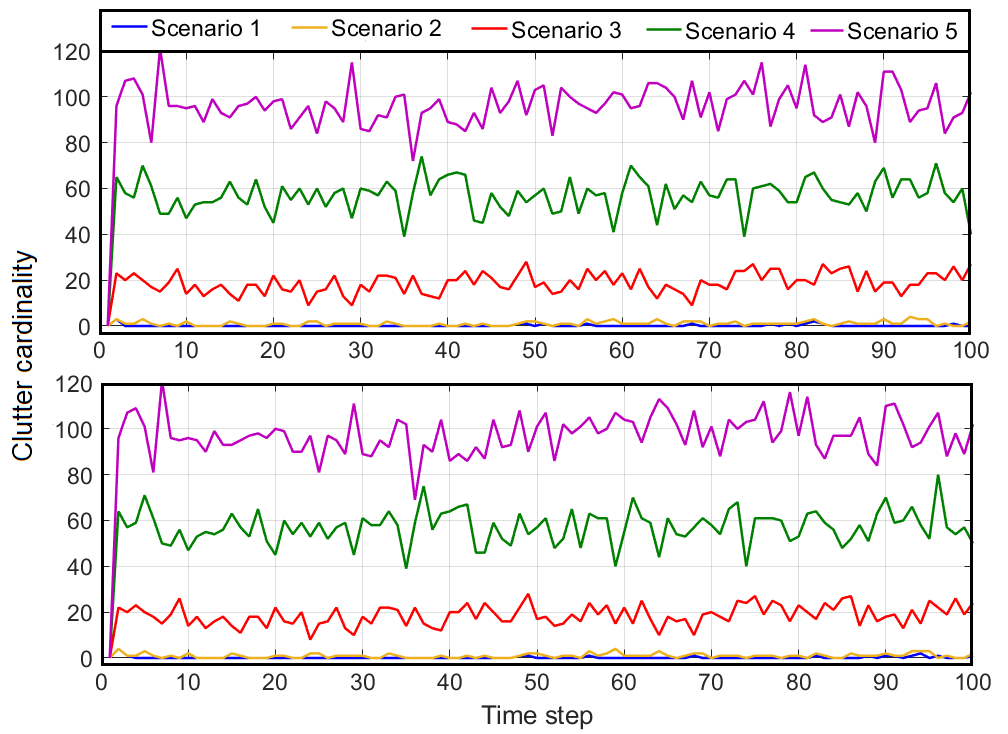}
\par\end{centering}
\caption{Estimated clutter cardinality in synthetic migration experiment (top:
PA, bottom: UA).\label{fig:exp2-clutter-synthetic}}
\end{figure}

\begin{figure*}
\begin{centering}
\includegraphics[width=0.9\textwidth]{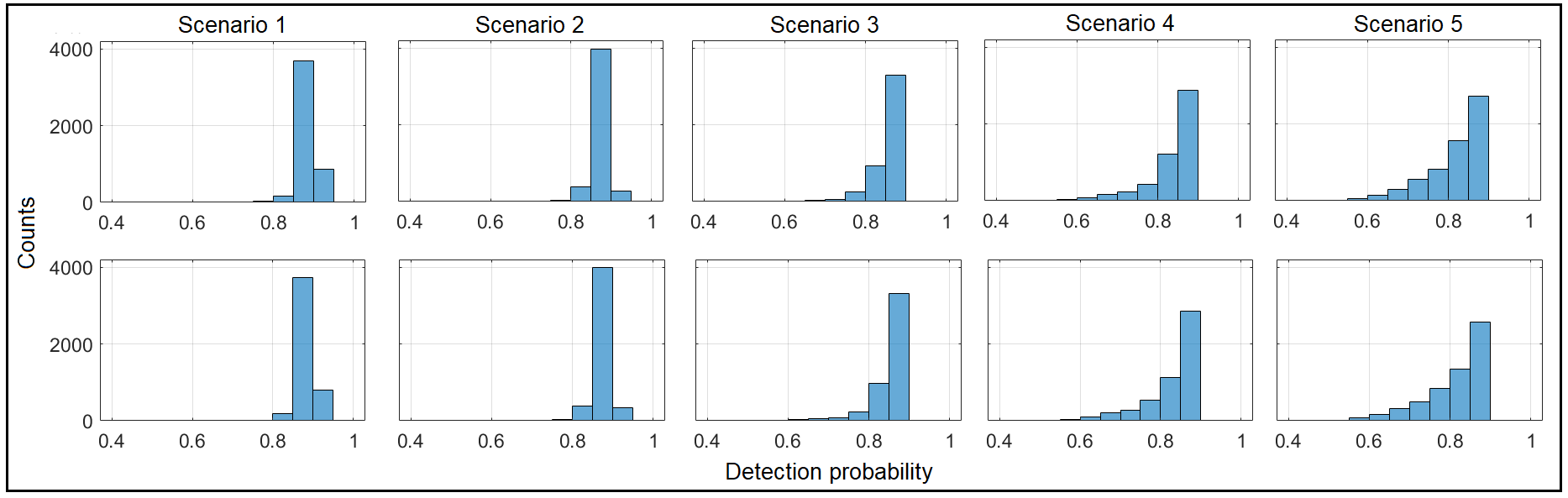}
\par\end{centering}
\caption{Histograms of cell detection probability (for all time steps) in synthetic
migration experiment (top: PA, bottom: UA).\label{fig:exp2-pd-histo-synthetic}}
\end{figure*}

Fig. \ref{fig:exp2-est-card-synthetic} shows that both PA and UA
accurately estimate the number of cells in scenarios 1 to 4 but exhibits
overestimation in scenario 5 due to the high clutter rate. Fig. \ref{fig:exp2-pd-synthetic}
shows that the average estimated detection probability decreases across
time from scenario 1 to 5. This is due to the difficulty in detection
when the cell density increases. The plots of clutter cardinality
in Fig. \ref{fig:exp2-clutter-synthetic} corroborate the average
false positive rates shown in Tab. \ref{fig:exp2-pd-synthetic}. To
provide further insight on the estimated detection probability, we
show the histograms of cell detection probability across all time
steps for all 5 scenarios in Fig. \ref{fig:exp2-pd-histo-synthetic}.
Note that the detection probability of a cell is dependent on its
state. Hence, more cells with low detection probability are estimated
across different scenarios as seen in Fig. \ref{fig:exp2-pd-histo-synthetic},
even though Fig. \ref{fig:exp2-pd-synthetic} does not show significant
reduction in average estimated detection probability. Tab. \ref{tab:exp2-ME}
shows PA has slightly higher mitotic error compared to UA. 

\begin{table}

\caption{The time-averaged error of estimated mitotic events counts.\label{tab:exp2-ME}}

\centering{}%
\begin{tabular}{|c|c|c|c|c|c|}
\hline 
\multicolumn{1}{|c|}{} & \textbf{\small{}Scen. 1} & \textbf{\small{}Scen. 2} & \textbf{\small{}Scen. 3} & \textbf{\small{}Scen. 4} & \textbf{\small{}Scen. 5}\tabularnewline
\hline 
\textbf{\small{}PA} & {\small{}1.62} & {\small{}1.53} & {\small{}1.47} & {\small{}1.40} & {\small{}1.23}\tabularnewline
\hline 
\textbf{\small{}UA} & {\small{}1.54} & {\small{}1.49} & {\small{}1.43} & {\small{}1.36} & {\small{}1.58}\tabularnewline
\hline 
\end{tabular}
\end{table}

We also provide a performance comparison with other state-of-the-art
methods: MHT \cite{CTP002} via icy\textsuperscript{\textcopyright}
\cite{software_icy} (icy-MHT), global optimization with Viterbi linking
algorithm \cite{CTP-D010} via BaxterAlgorithm \cite{T006} (Viterbi
Linking), JPDA with Interacting Multiple Models (IMM-JPDA) \cite{066},
and Linear Assignment Problem (LAP) via CellProfiler\textsuperscript{\textcopyright}
\cite{CTP045} (CellProfiler-LAP). The motion model parameters are
the same as per our proposed algorithm. Other parameters are taken
from actual values in Tab. \ref{tab:exp2-pD-clutter} or tuned via
the parameters estimation routines provided with the software, prior
to tracking.

\begin{comment}
detection probability and clutter rate are individually tuned so that
these state-of-the-art algorithms achieve good tracking performance
for each scenario.
\end{comment}

\begin{figure}
\begin{centering}
\includegraphics[width=0.45\textwidth]{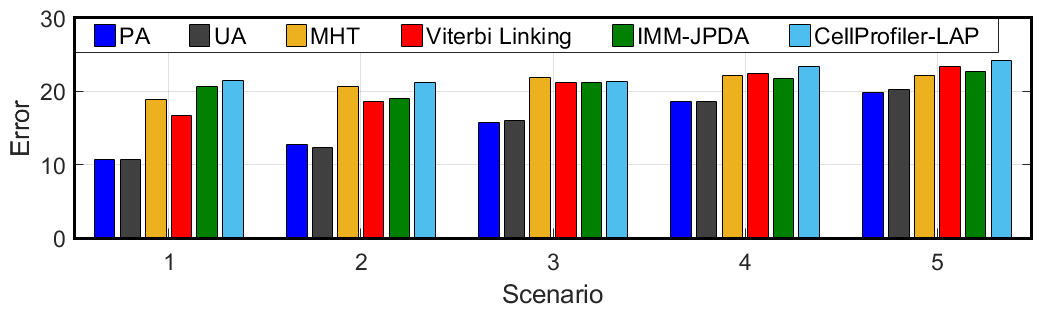}
\par\end{centering}
\caption{$\textrm{OSPA}^{\textrm{(2)}}$ error (lower is better) of different
cell tracking methods across all scenarios in synthetic migration
experiment (evaluated over the entire tracking period).\label{fig:exp2-OSPA2-method-synthetic}}
\end{figure}

The $\textrm{OSPA}^{\textrm{(2)}}$ errors over the entire scenario
for all filters under consideration are plotted in Fig. \ref{fig:exp2-OSPA2-method-synthetic}.
It can be seen that the proposed GLMB-based methods have the lowest
error, keeping in mind that they have no knowledge of the false positive
and negative rates. \textcolor{black}{The errors for the other algorithms
are similar in all scenarios, and is highest for CellProfiler-LAP
in scenarios 1, 2, 4 and 5, and highest for icy-MHT in scenario 3.
The true number of distinct tracks in this experiment is 357. Icy-MHT
and Viterbi Linking algorithms overestimate the number of tracks while
IMM-JPDA and Cellprofilter-LAP underestimate. On the other hand, our
methods yield the lowest error in the number of tracks across all
scenarios.}

In addition to the $\textrm{OSPA}^{\textrm{(2)}}$ metric, we report
the TRA score \cite{CTP019} to evaluate lineage estimation (IMM-JPDA
is excluded as it does not provide lineage). Instead of computing
the overlap region between the true and the computed cells, we use
the Euclidean distance between them as the matching cost. Matches
that have distances lower than 25 pixels are counted as true positives.
The standard computation for TRA scores then gives the false positives,
false negatives links etc. Equal weights are used for all 5 types
of errors (merged, false negatives, false positives, false negative
links, false positive links and incorrect semantic links). Readers
are referred to \cite{CTP019} for more details on TRA score. 

The results in Tab. \ref{tab:exp2-tra-score-synthetic} indicate that
PA has the best performance in TRA score on all scenarios while UA
is the second best in scenarios 1 to 4. MHT performs poorly in the
3 most challenging scenarios with a high amount of false tracks which
is presumably due to the high clutter rate. 

\begin{table}
\caption{TRA scores (higher is better) for different algorithms.\label{tab:exp2-tra-score-synthetic}}

\begin{centering}
\begin{tabular}{|c|c|c|c|c|c|}
\hline 
\multicolumn{1}{|c|}{} & \textbf{\footnotesize{}Scen. 1} & \textbf{\footnotesize{}Scen. 2} & \textbf{\footnotesize{}Scen. 3} & \textbf{\footnotesize{}Scen. 4} & \textbf{\footnotesize{}Scen. 5}\tabularnewline
\hline 
\textbf{\footnotesize{}PA} & \textbf{\footnotesize{}0.7458} & \textbf{\footnotesize{}0.7429} & \textbf{\footnotesize{}0.6319} & \textbf{\footnotesize{}0.3295} & \textbf{\footnotesize{}0.1413}\tabularnewline
\hline 
\textbf{\footnotesize{}UA} & {\footnotesize{}0.7447} & {\footnotesize{}0.7331} & {\footnotesize{}0.5812} & {\footnotesize{}0.3138} & {\footnotesize{}0.0678}\tabularnewline
\hline 
\textbf{\footnotesize{}icy-MHT} & {\footnotesize{}0.5584} & {\footnotesize{}0.1675} & {\footnotesize{}0} & {\footnotesize{}0} & {\footnotesize{}0}\tabularnewline
\hline 
\textbf{\footnotesize{}Viterbi Linking} & {\footnotesize{}0.5789} & {\footnotesize{}0.4763} & {\footnotesize{}0.2273} & {\footnotesize{}0.1575} & {\footnotesize{}0.0811}\tabularnewline
\hline 
\textbf{\footnotesize{}\negthinspace{}\negthinspace{}CellProfiler-LAP\negthinspace{}\negthinspace{}} & {\footnotesize{}0.4821} & {\footnotesize{}0.4694} & {\footnotesize{}0.4094} & {\footnotesize{}0.0044} & {\footnotesize{}0}\tabularnewline
\hline 
\end{tabular}
\par\end{centering}
\begin{comment}
\begin{center}
\begin{tabular}{|c|c|c|c|c|c|}
\hline 
\multicolumn{1}{|c|}{} & \textbf{Scenario 1} & \textbf{Scenario 2} & \textbf{Scenario 3} & \textbf{Scenario 4} & \textbf{Scenario 5}\tabularnewline
\hline 
\textbf{Our method} & \textbf{0.5854} & \textbf{0.7158} & \textbf{0.5197} & \textbf{0.2539} & \textbf{0.1312}\tabularnewline
\hline 
\textbf{icy-MHT} & 0.5584 & 0.1675 & 0 & 0 & 0\tabularnewline
\hline 
\textbf{Viterbi Linking} & 0.5789 & 0.4763 & 0.2273 & 0.1575 & 0.0811\tabularnewline
\hline 
\textbf{IMM-JPDA} & \textbf{0.6667} & 0.6458 & 0.4983 & \textbf{0.2628} & 0.0077\tabularnewline
\hline 
\textbf{CellProfiler-LAP} & 0.4821 & 0.4694 & 0.4094 & 0.0044 & 0\tabularnewline
\hline 
\end{tabular}
\par\end{center}
\end{comment}
\end{table}
We compare the computation times of PA and UA (on a 12-core machine
at 1.5 GHz with parallelization applied where possible) in Tab. \ref{tab:exp2-comp-time}
\textcolor{black}{noting that both approximate filters exhibit similar
performance. The difference in computation times for PA and UA is
not significant in the first 2 scenarios but the gaps noticeably widened
from scenario 3 onward. Other methods took roughly 5 to 15 minutes
to compute the sequences. The additional computation times for the
proposed methods is due to the fact it jointly estimates the de}tection
probability and clutter rate whereas the existing methods require
these parameters to be known prior to tracking. While these execution
times are only meant to be indicative, they suggest that all methods
are suitable for live image experiments, which typically have observation
intervals of 10-15 minutes. 

\begin{comment}
\begin{table}
\caption{{[}CHANGE TO BAR PLOT, ADD AVERAGE OSPA2{]} Number of distinct cells
tracks from different cell tracking methods in synthetic migration
experiment. \label{tab:exp2-num-traj-synthetic}}

\centering{}%
\begin{tabular}{|c|c|c|c|c|c|}
\hline 
\multicolumn{1}{|c|}{} & \textbf{Scen. 1} & \textbf{Scen. 2} & \textbf{Scen. 3} & \textbf{Scen. 4} & \textbf{Scen. 5}\tabularnewline
\hline 
\textbf{Truth} & \multicolumn{5}{c|}{357}\tabularnewline
\hline 
\textbf{Our method} & \textbf{314} & 279 & \textbf{268} & \textbf{294} & \textbf{331}\tabularnewline
\hline 
\textbf{icy-MHT} & 592 & 950 & 1381 & 1512 & 1568\tabularnewline
\hline 
\textbf{Viterbi Linking} & 410 & \textbf{417} & 517 & 634 & 925\tabularnewline
\hline 
\textbf{IMM-JPDA} & 121 & 175 & 166 & 288 & 419\tabularnewline
\hline 
\textbf{CellProfiler-LAP} & 118 & 135 & 219 & 721 & 1318\tabularnewline
\hline 
\end{tabular}
\end{table}
\end{comment}

\begin{table}
\caption{Computation times of proposed filters across different scenarios.\label{tab:exp2-comp-time}}

\centering{}%
\begin{tabular}{|c|c|c|c|c|c|}
\hline 
\multicolumn{1}{|c|}{} & \textbf{\small{}Scen. 1} & \textbf{\small{}Scen. 2} & \textbf{\small{}Scen. 3} & \textbf{\small{}Scen. 4} & \textbf{\small{}Scen. 5}\tabularnewline
\hline 
\textbf{\small{}PA (min)} & {\small{}20} & {\small{}18} & {\small{}22} & {\small{}38} & {\small{}66}\tabularnewline
\hline 
\textbf{\small{}UA (min)} & {\small{}24} & {\small{}25} & {\small{}89} & {\small{}131} & {\small{}187}\tabularnewline
\hline 
\end{tabular}
\end{table}
\begin{comment}
\begin{figure}
\begin{centering}
\includegraphics[width=0.45\textwidth,bb = 0 0 200 100, draft, type=eps]{../../../../../AppData/Local/Microsoft/Windows/INetCache/Content.Outlook/D41VABBD/graphic/exp2/comp_time.png}
\par\end{centering}
\caption{Computation time in seconds for our method with the given settings
in Tab. \ref{tab:tracker-parameter-sim}.\label{fig:exp2-comp-time-synthetic}}
\end{figure}
\end{comment}

\subsection{Breast Cancer Cells Migration Sequence\label{subsec:Real-breast-cancer}}

This experiment considers a real breast cancer cell (MDA-MB-231) dataset
in an 88 frame migration sequence of time-lapsed images taken every
15 minutes by inverted microscope. Cell detections are extracted from
the images via a neural network called FRCNN-ResNet101 \cite{M-FRCNN}
and trained on a separate set of images of the same type of cell.
Since mitotic cells have different features to normal cells, we use
the output of the last fully-connected layer of FRCNN-ResNet101 to
capture their features (1000-dimensional vectors). Specifically, we
feed the training sub-images (bounding boxes) of mitotic and normal
cells to FRCNN-ResNet101, and then extract the feature vectors from
the last fully-connected layer to form $\mathbb{F}_{1}$ (the set
of feature vectors of normal cells) and $\mathbb{F}_{2}$ (the set
of feature vectors of mitotic cells). To obtain each appearance measurement,
we extract the centroid of the box (used as the detected location
$\varrho$ of the cell\footnote{Although the detector returns bounding boxes, in this experiment,
using their centroids is sufficient for estimating cell migration
patterns and lineages. Further, since cells are relatively small compared
to the image size, this approach strikes a balance between performance
and computational cost.}), and then feed the sub-image of the detected cell to FRCNN-ResNet101
to generate $\alpha$.

\begin{figure}
\begin{centering}
\includegraphics[width=0.45\textwidth]{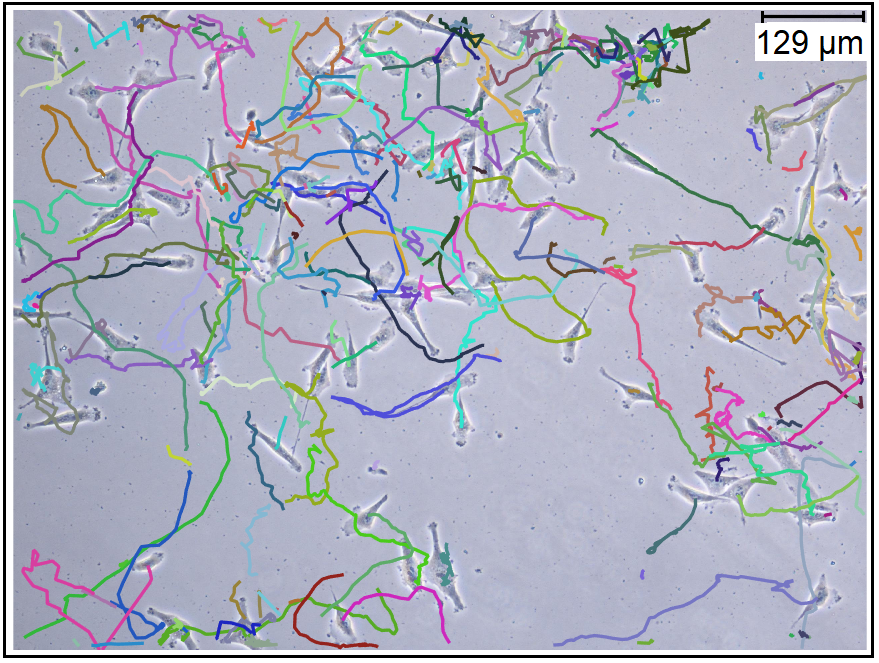}
\par\end{centering}
\caption{Estimated breast cancer cell trajectories.\label{fig:exp3est-cell-traj}}
\end{figure}
\begin{figure}
\begin{centering}
\includegraphics[width=0.3\textwidth]{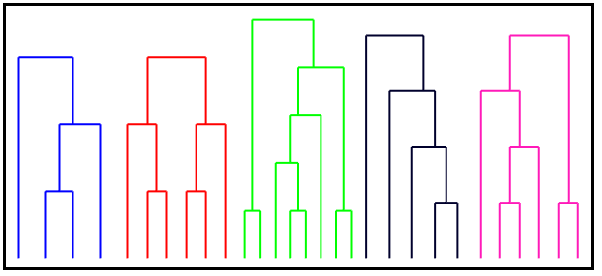}
\par\end{centering}
\caption{Estimated cell lineages of 5 largest breast cancer cell families,
each distinct color indicates a cell family.\label{fig:exp3-est-cell-lineage}}
\end{figure}
\begin{figure}
\begin{centering}
\includegraphics[width=0.45\textwidth]{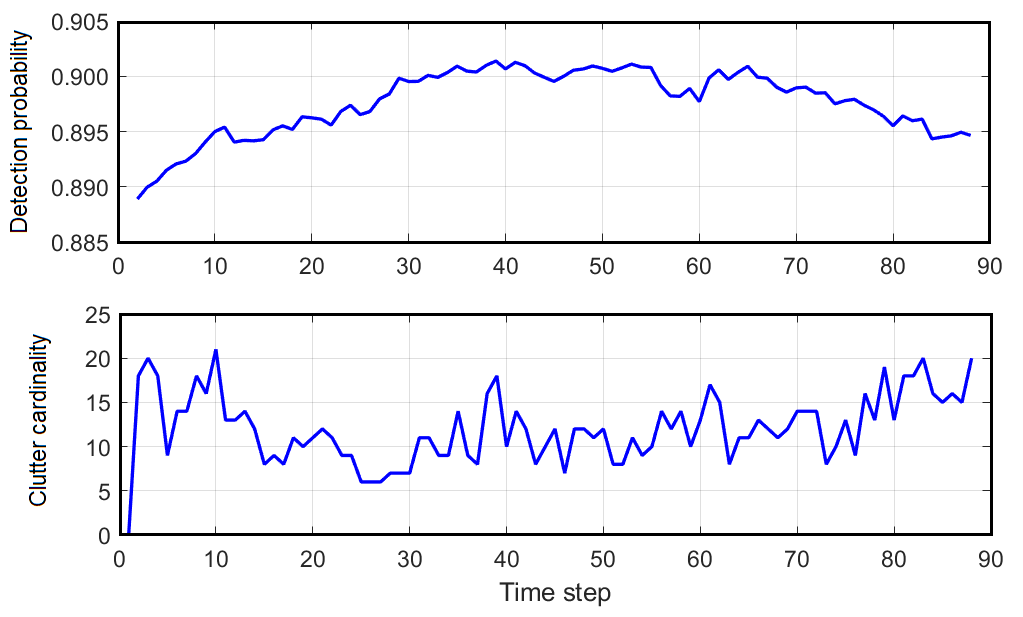}
\par\end{centering}
\caption{Average estimated detection probability (top) and clutter cardinality
(bottom) for breast cancer cells.\label{fig:exp3est-pd-clutter}}
\end{figure}
Due to the more challenging setting of this experiment, i.e. relatively
large number of cells with high uncertainty in dynamic and measurements,
we only apply the proposed PA filter. The same dynamic and observation
model parameters as in the synthetic data experiments are used, except
for the mode likelihood, which is given by $g^{(a)}(\alpha|i)\!\!=\sum_{F\in\mathbb{F}_{i}}\left\Vert \alpha-F\right\Vert /\left|\mathbb{F}_{i}\right|$for
$i=1,2.$ The filter parameters are also the same as in the synthetic
data experiments, except that the number of maximum components is
increased to 50000. Fig. \ref{fig:exp3est-cell-traj} shows the trajectories
of the estimated cells at the end of the sequence. Fig. \ref{fig:exp3-est-cell-lineage}
also illustrates the estimated cell lineage of the 5 largest cell
families. The results demonstrate the capability of the proposed algorithm
in tracing cells and their lineages over long periods of time. Fig.
\ref{fig:exp3est-pd-clutter} shows the average estimated detection
probability and clutter cardinality across different time steps.

\begin{figure}
\begin{centering}
\includegraphics[width=0.45\textwidth]{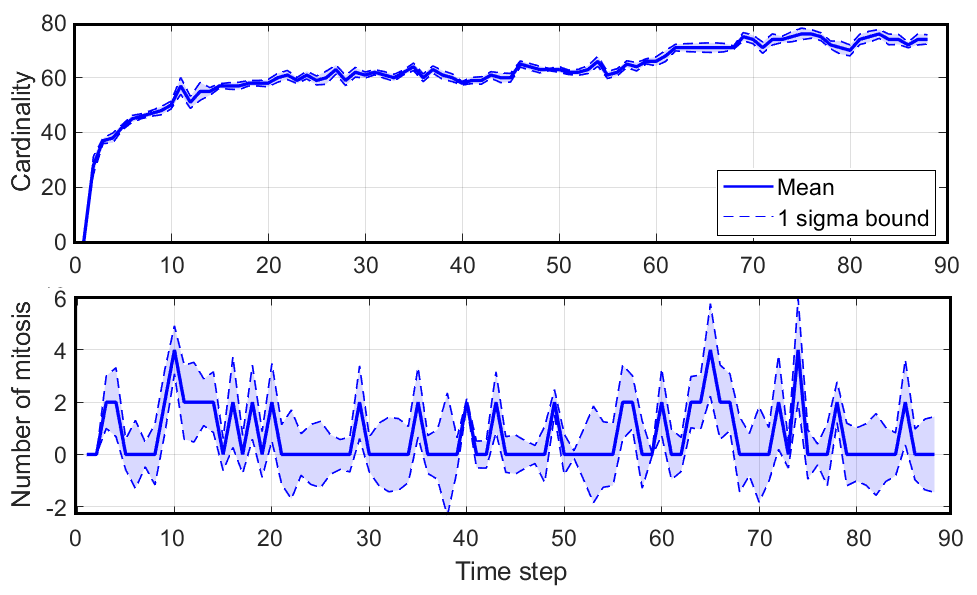}
\par\end{centering}
\caption{Cardinality and mitotic events counts for breast cancer cells.\label{fig:exp3-cell-card-stats}}
\end{figure}

\begin{figure}
\begin{centering}
\includegraphics[width=0.45\textwidth]{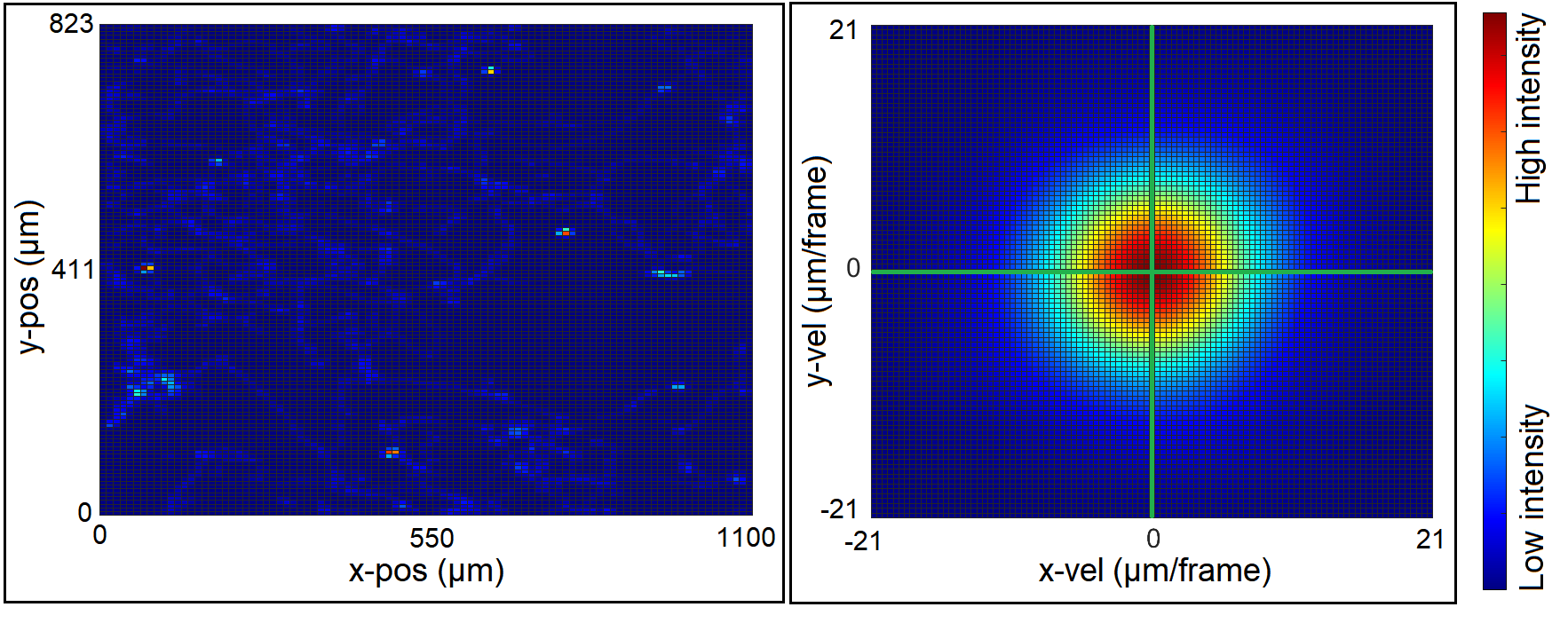}
\par\end{centering}
\caption{Cell intensity or concentration in position space and velocity space
for breast cancer cells.\label{fig:exp3-pos-heatmap}}
\end{figure}

From the posterior statistics on the population size and mitotic events
in Fig. \ref{fig:exp3-cell-card-stats}, observe that the number of
cells increases from approximately 30 in the first frame to nearly
80 in the last frames. Moreover, the relatively tight 1-sigma bounds
suggest that the algorithm has high confidence on the estimated cardinality
statistics. \textcolor{black}{Fig. \ref{fig:exp3-pos-heatmap} shows
heat maps of cell concentration in the position and velocity spaces,
computed by averaging the intensity function (or probability hypothesis
density) over the entire period.} Such plots can provide valuable
insights into the cell population, such as the regions where cells
are more likely to exist (bright spots), or the overall directional
drift of cells (in this case slightly downward). Interestingly, the
observed cell concentration in the velocity space has a distinctly
Gaussian profile. 


\section{Conclusions\label{sec:Conclusions}}

We have proposed labeled RFS solutions for tracking cell trajectories
and their lineages. These solutions are based on: a spawning model
that takes into account cell lineage and changes in cell appearance
prior to division; and multi-object tracking filters for this spawning
model. Additionally, the proposed solutions offer the tools for characterizing
uncertainty on inferred results, and operate in environments with
unknown detection probability and the clutter rate, which is invariably
the case in cell experiments. The numerical case studies demonstrate
the capability of the proposed solutions to reliably estimate the
cell tracks and their lineages. The synthetic data case study, with
different level of tracking difficulty, demonstrates significant improvements
over existing methods. The real data case study with breast cancer
cells migration illustrates the capability to characterize uncertainty
on inferred results in providing insightful statistics on the migration
of the cell population. 

\section{Appendix\label{sec:Appendix}}

\subsection{Proof of Proposition \ref{prop:exact_prediction} \label{subsec:Proof-of-pred-aprx-1}}

Substuting (\ref{eq:exact_transition}) to (\ref{e:MTBF}) yields
\begin{equation}
{\textstyle \boldsymbol{\pi}_{+}\!\left(\boldsymbol{X}_{+}\right)=\boldsymbol{f}_{B,+}(\boldsymbol{B})\int}\boldsymbol{f}_{G,+}(\boldsymbol{Y}|\boldsymbol{X})\boldsymbol{\pi}\left(\boldsymbol{X}\right)\delta\boldsymbol{X},\label{eq:app-pred-density}
\end{equation}
where $\boldsymbol{Y}$ =$\boldsymbol{X}_{+}-(\mathbb{X}\times\mathbb{B}_{+})$
and $\boldsymbol{B}$ =$\boldsymbol{X}_{+}\cap(\mathbb{X}\times\mathbb{B}_{+})$.

The integral term can be written as \allowdisplaybreaks
\begin{eqnarray*}
\!\!\!\!\!\!\!\!\!\!\!\!\!\! &  & \!\!\!\!\!\!\!\!\!\!\!\!\!\!\!\!\int\Delta(\boldsymbol{X})\sum_{I,\xi}\!\omega^{(I,\xi)}\delta_{I}[\mathcal{L}(\boldsymbol{X})]p^{(\xi)}(\boldsymbol{X})\boldsymbol{f}_{+}^{(G)}(\boldsymbol{Y}|\boldsymbol{X})\delta\boldsymbol{X}\\
\!\!\!\!\! & = & \!\!\!\!\Delta(\boldsymbol{Y})\sum_{L\subseteq\mathbb{L}}\sum_{I,\xi}\omega^{(I,\xi)}\delta_{I}[L]\eta_{G,+}^{(L,\xi)}(\mathcal{L}(\boldsymbol{Y}))p_{G,+}^{(L,\xi)}(\boldsymbol{Y})\\
\!\!\!\!\! & = & \!\!\!\!\Delta(\boldsymbol{Y})\sum_{I,\xi}\!\omega^{(I,\xi)}\eta_{G,+}^{(I,\xi)}(\mathcal{L}(\boldsymbol{Y}))p_{G,+}^{(I,\xi)}(\boldsymbol{Y}).
\end{eqnarray*}
The second line follows from Lemma 7 of \cite{017}. The sum over
$L$ vanishes for $L\neq I$, yielding the last line. Further, we
replace $\eta_{G,+}^{(I,\xi)}(L)$ by $\sum_{J\subseteq\mathbb{G}_{+}(\mathbb{L})}\delta_{J}(L)\eta_{G,+}^{(I,\xi)}(J)$,
and noting that the weights of the LMB birth model (\ref{eq:LMB_birth})
can be rewritten as $w_{B,+}(L)=\sum_{J'\subseteq\mathbb{B}_{+}}\delta_{J'}(L)w_{B,+}(J')$,
(\ref{eq:app-pred-density}) becomes
\begin{eqnarray*}
\boldsymbol{\pi}_{+}(\boldsymbol{X}_{+})\!\!\!\!\! & = & \!\!\!\!\Delta(\boldsymbol{Y})\Delta(\boldsymbol{B})\!\sum_{I,\xi}\!\sum_{J\subseteq\mathbb{G}_{+}(\mathbb{L})}\omega^{(I,\xi)}\delta_{J}[\mathcal{L}(\boldsymbol{Y})]\eta_{G,+}^{(I,\xi)}(J)\\
\!\!\!\!\! &  & \!\!\!\!\times p_{G,+}^{(I,\xi)}(\boldsymbol{Y})\sum_{J'\subseteq\mathbb{B}_{+}}\delta_{J'}[\mathcal{L}(\boldsymbol{B})]w_{B,+}(J')[p_{B,+}]^{\boldsymbol{B}}\\
\!\!\!\!\! & = & \!\!\!\!\Delta(\boldsymbol{X}_{+})\sum_{I,\xi}\sum_{(J\uplus J')\subseteq(\mathbb{G}_{+}(\mathbb{L})\uplus\mathbb{B}_{+})}\!\!\delta_{J\uplus J'}[\mathcal{L}(\boldsymbol{Y})\uplus\mathcal{L}(\boldsymbol{B})]\\
\!\!\!\!\! &  & \!\!\!\!\times\omega^{(I,\xi)}w_{B,+}(J')\eta_{G,+}^{(I,\xi)}(J)[p_{B,+}]^{\boldsymbol{B}}p_{G,+}^{(I,\xi)}(\boldsymbol{Y})\\
\!\!\!\!\! & = & \!\!\!\!\Delta(\boldsymbol{X}_{+})\sum_{I,\xi,I_{+}}\omega^{(I,\xi)}w_{B,+}(I_{+}\cap\mathbb{B}_{+})\times\\
\!\!\!\!\! &  & \!\!\!\!\eta_{G,+}^{(I,\xi)}(I_{+}-\mathbb{B}_{+})\delta_{I_{+}}[\mathcal{L}(\boldsymbol{X}_{\!+})][p_{B,+}]^{\boldsymbol{B}}p_{G,+}^{(I,\xi)}(\boldsymbol{Y}).
\end{eqnarray*}
The second equation follows from $\boldsymbol{X}_{+}=\boldsymbol{Y}\uplus\boldsymbol{B}$
and $\Delta(\boldsymbol{X}_{+})=\Delta(\boldsymbol{Y})\Delta(\boldsymbol{B})$.
The last equation is obtained by substituting $I_{+}$= $J\uplus J'$,
and  $\mathbb{L}_{+}=\mathbb{G}_{+}(\mathbb{L})\uplus\mathbb{B}_{+}$.
$\boldsymbol{\enskip\blacksquare}$

\subsection{Proof of Proposition \ref{prop:exact_update} \label{subsec:Proof-of-upd-aprx}}

Substituting $\boldsymbol{Y}=\boldsymbol{X}_{+}-\mathbb{X}\times\mathbb{B}_{+}$
and $\boldsymbol{B}=\boldsymbol{X}_{+}\cap\mathbb{X}\times\mathbb{B}_{+}$
into (\ref{eq:GLMB_apprx_pred}) gives
\begin{eqnarray*}
\boldsymbol{\pi}_{\!+}\!(\boldsymbol{X}_{\!+})\!\!\!\!\! & = & \!\!\!\!\Delta(\boldsymbol{X}_{\!+})\!\!\sum_{I,\xi,I_{+}}\!\!\omega_{+}^{(I,\xi,I_{+\!})}\delta_{I_{+\!}}[\mathcal{L}(\boldsymbol{X}_{\!+})]\left[p_{B,+}\right]{}^{\boldsymbol{B}}p_{G,+}^{(I,\xi)}(\boldsymbol{Y}).
\end{eqnarray*}
Applying Bayes rule, we have
\begin{eqnarray*}
\boldsymbol{\pi}_{\!+}\!(\boldsymbol{X}_{\!+}|Z_{+})\!\!\!\! & \propto & \!\!\!\!\Delta(\boldsymbol{X}_{\!+})\!\!\sum_{I,\xi,I_{+}}\!\!\omega_{+}^{(I,\xi,I_{+\!})}\delta_{I_{+\!}}[\mathcal{L}(\boldsymbol{X}_{\!+})]\left[p_{B,+}\right]{}^{\boldsymbol{B}}\\
 &  & \!\!\!\!p_{G,+}^{(I,\xi)}(\boldsymbol{Y})\sum_{\theta_{+}\in\Theta_{+}}1_{\Theta_{+}(\mathcal{L}(\boldsymbol{X}_{+}))}(\theta_{+})[\varPsi_{Z_{+}}^{(\theta_{+})}]^{\boldsymbol{X}_{+}}.
\end{eqnarray*}
Further, substituting $[\varPsi_{Z_{+}}^{(\theta_{+})}]^{\boldsymbol{X}_{+}}$
= $[\varPsi_{Z_{+}}^{(\theta_{+})}]^{\boldsymbol{B}}[\varPsi_{Z_{+}}^{(\theta_{+})}]^{\boldsymbol{Y}}$
and $I_{+}=\mathcal{L}(\boldsymbol{X}_{+})$ yields
\begin{multline*}
\boldsymbol{\pi}_{\!+}\!(\boldsymbol{X}_{\!+}|Z_{+})\propto\Delta(\boldsymbol{X}_{\!+})\!\!\sum_{I,\xi,I_{+},\theta_{+}}\!\omega_{+}^{(I,\xi,I_{+\!})}\delta_{I_{+\!}}[\mathcal{L}(\boldsymbol{X}_{\!+})]\!\times\\
1_{\Theta_{+}(I_{+})}(\theta_{+})\left[p_{B,+}\varPsi_{Z_{+}}^{(\theta_{+})}\right]{}^{\boldsymbol{B}}p_{G,+}^{(I,\xi)}(\boldsymbol{Y})[\varPsi_{Z_{+}}^{(\theta_{+})}]^{\boldsymbol{Y}}\\
=\Delta(\boldsymbol{X}_{\!+})\!\!\sum_{I,\xi,I_{+},\theta_{+}}\!\!\omega_{+}^{(I,\xi,I_{+\!})}1_{\Theta_{+}(I_{+})}(\theta_{+})\delta_{I_{+\!}}[\mathcal{L}(\boldsymbol{X}_{\!+})]\times\\
\!\left[\eta_{B,Z_{+}}^{(\theta_{+})}\right]^{\mathcal{L}(\boldsymbol{B})}\eta_{G,Z_{+}}^{(I,\xi,\theta_{+})}(\mathcal{L}(\boldsymbol{Y}))\left[p_{B,Z_{+}}^{(\theta_{+})}\!\right]\!^{\!\boldsymbol{B}}p_{G,Z_{+}}^{(I,\xi,\theta_{+})}(\boldsymbol{Y}).\enskip\blacksquare
\end{multline*}

\subsection{Equivalence of (\ref{eq:pred_aprx_weight}) and (\ref{eq:pred_aprx_weight_equivalence})
\label{subsec:Equivalence-of-apprx-pred-weight}}

Expanding (\ref{eq:pred_aprx_weight}) we have
\begin{multline*}
\tilde{\omega}{}_{Z_{+}}^{(I,\xi,I_{+},\theta_{+})}=\omega^{(I,\xi)}w_{B,+}(I_{+}\cap\mathbb{B}_{+})\eta_{G,+}^{(I,\xi))}(I_{+}-\mathbb{B}_{+})\times\\
1_{\Theta_{+}(I_{+})}(\theta_{+})\left[\eta_{B,Z_{+}}^{(\theta_{+})}\right]^{I_{+}\cap\mathbb{B}_{+}}\left[\tilde{\eta}_{G,Z_{+}}^{(I,\xi,I_{+},\theta_{+})}\right]^{(I_{+}-\mathbb{B}_{+})}.
\end{multline*}
For the birth terms, we write
\begin{eqnarray*}
w_{B,+}(I_{+}\cap\mathbb{B}_{+}) & \!\!\!\!=\!\!\!\! & [1-r_{B,+}(\cdot)]^{\mathbb{B}_{+}-I_{+}}[r_{B,+}(\cdot)]^{I_{+}\cap\mathbb{B}_{+}},\\
\left[\eta_{B,Z_{+}}^{(\theta_{+})}\right]^{I_{+}\cap\mathbb{B}_{+}} & \!\!\!\!=\!\!\!\! & \left[\int p_{+}^{(B)}(x_{+},\cdot)\psi_{Z_{+}}^{(\theta_{+}(\cdot))}(x_{+},\cdot)dx_{+}\right]^{I_{+}\cap\mathbb{B}_{+}}.
\end{eqnarray*}
For the survival terms we write 
\begin{align*}
\eta_{G,+}^{(I,\xi)}(I_{+}-\mathbb{B}_{+})\!=\! & \left\langle \prod_{\ell\in I}q_{G,+}^{(\ell,\xi)}((I_{+}-\mathbb{B}_{+})\!\cap\!\mathbb{G}_{+}(\ell))\right\rangle \!(I_{+}-\mathbb{B}_{+})\\
=\! & \prod_{\ell\in I}\left\langle q_{G,+}^{(\ell,\xi)}(I_{+}\cap\mathbb{G}_{+}(\ell))\right\rangle (I_{+}\cap\mathbb{G}_{+}(\ell))\\
=\! & \prod_{\ell\in I}\eta_{G,+}^{(\ell,\xi)}(I_{+}\cap\mathbb{G}_{+}(\ell)),
\end{align*}
where the second equation follows from the separation of independent
variables for integration, and
\begin{multline*}
\!\!\!\!\!\!\!\![\tilde{\eta}_{G,Z_{+}}^{(I,\xi,I_{+},\theta_{+})}]^{(I_{+}-\mathbb{B}_{+})}\!=\!\!\!\!\prod_{u\in I_{+}-\mathbb{B}_{+}}\left\langle p_{G,+}^{(I,\xi,I_{+}-\mathbb{B}_{+})}(\cdot,u),\varPsi_{Z_{+}}^{(\theta_{+})}(\cdot,u)\right\rangle \\
=\prod_{\ell\in I}\prod_{\ell_{+}\in I_{+}\cap\mathbb{G}_{+}(\ell)}\left\langle p_{G,+}^{(\ell,\xi,I_{+}\cap\mathbb{G}_{+}(\ell))}(\cdot,\ell_{+}),\varPsi_{Z_{+}}^{(\theta_{+})}(\cdot,\ell_{+})\right\rangle .
\end{multline*}
Grouping the appropriate terms yields (\ref{eq:pred_aprx_weight_equivalence}).$\enskip\blacksquare$

\section{Proof of Proposition 6\label{sec:Proof-of-Proposition-6}}

It suffices to show that\textit{
\begin{align}
\pi_{n}(\gamma_{n}|\gamma_{\bar{n}}) & \propto\lambda_{n}^{(I,\xi)}(\gamma_{n})\prod_{i\in\bar{n}}\Upsilon_{\{1:M\}}(\gamma_{n},\gamma_{i}),\label{eq:Gibbs-n-proposal-equi-1}
\end{align}
}where
\[
\Upsilon_{\{1:M\}}(\gamma_{n},\gamma_{i})=\prod_{(q,r)=(1,1)}^{(3,3)}(1-1_{\{1:M\}}(\gamma_{n,q})\delta_{\gamma_{n,q}}[\gamma_{i,r}]).
\]
 To prove (\ref{eq:Gibbs-n-proposal-equi-1}), we first show that
$1_{\Gamma}(\gamma)$ can be written in the following product form
\begin{align}
1_{\Gamma}(\gamma) & =1_{\Gamma(\bar{n})}(\gamma_{\bar{n}})\prod_{i\in\bar{n}}\Upsilon_{\{1:M\}}(\gamma_{n},\gamma_{i}),\label{eq:gamma_inclusion_factor-1}
\end{align}
where $\Gamma(\bar{n})$ is the set of all 1-1 positive $\gamma_{\bar{n}}$.

For $n\in\{1:P\}$, since $\gamma_{n}\in\mathbb{D}_{+}\uplus\mathbb{N}_{+}$
the positive 1-1 condition is guaranteed by the definition of $\lambda_{n}^{(I,\xi)}$.
It remains to prove that if $\prod_{i\in\bar{n}}\Upsilon_{\{1:M\}}(\gamma_{n},\gamma_{i})=1$
then $\gamma$ is positive 1-1. 

To prove (\ref{eq:gamma_inclusion_factor-1}), we will show that:
(a) if $\gamma$ is positive 1-1 then RHS of (\ref{eq:gamma_inclusion_factor-1})
equates to 1; and (b) if $\gamma$ is not positive 1-1 then RHS of
(\ref{eq:gamma_inclusion_factor-1}) equates to 0.

For (a), assume that $\gamma$ is positive 1-1, then for any $i\neq j\in\bar{n}$,
$\Upsilon_{\{1:M\}}(\gamma_{j},\gamma_{i})=1$ hence RHS equates to
1. For (b), assume that $\gamma$ is not positive 1-1, if $\gamma_{\bar{n}}$
is also not positive 1-1, hence there exists $i\neq j\in\bar{n}$
such that $\Upsilon_{\{1:M\}}(\gamma_{j},\gamma_{i})=0$ hence $1_{\Gamma(\bar{n})}(\gamma_{\bar{n}})=0$,
hence RHS equates to 0. If $\gamma_{\bar{n}}$ is positive 1-1, as
$\gamma$ is not positive 1-1, hence there exists $i$ and $j$ such
that $\Upsilon_{\{1:M\}}(\gamma_{j},\gamma_{i})=0$. Either $i$ or
$j$ has to equal $n$ as the 1-1 positive of $\gamma_{\bar{n}}$
are in $\bar{n}$ then we have a contradiction. Hence the RHS must
equate to 0. To prove (\ref{eq:Gibbs-n-proposal-equi-1}), as we are
interested in the relationship between $\pi_{n}(\gamma_{n}|\gamma_{\bar{n}})$
and $\gamma_{n}$, we can write
\[
\pi_{n}(\gamma_{n}|\gamma_{\bar{n}})\propto\pi(\gamma)\propto1_{\Gamma}(\gamma)\prod_{j=1}^{P}\lambda_{i}^{(I,\xi)}(\gamma_{j}).
\]
Applying (\ref{eq:gamma_inclusion_factor-1}) to the above expression
yields
\begin{multline*}
\pi_{n}(\gamma_{n}|\gamma_{\bar{n}})\\
\propto\lambda_{n}^{(I,\xi)}(\gamma_{n})\prod_{i\in\bar{n}}\Upsilon_{\{1:M\}}(\gamma_{n},\gamma_{i})1_{\Gamma(\bar{n})}(\gamma_{\bar{n}})\prod_{j\in\bar{n}}\lambda_{j}^{(I,\xi)}(\gamma_{j})\\
\propto\lambda_{n}^{(I,\xi)}(\gamma_{n})\prod_{i\in\bar{n}}\Upsilon_{\{1:M\}}(\gamma_{n},\gamma_{i}).\enskip\blacksquare
\end{multline*}

\section{Proof of Proposition 7\label{sec:Proof-of-Proposition-7}}

Note from the $n^{th}$ conditional given by (\ref{eq:Gibbs-n-proposal-equi-1})
that $\Upsilon_{\{1:M\}}(\gamma_{n},\gamma_{j})=1$, for each $j\in\{1:n-1\}$.
Hence, 
\begin{align*}
\!\!\pi_{n}(\gamma'_{n}|\gamma'_{1:n-1},\gamma{}_{n+1:P}) & =\lambda_{n}^{(I,\xi)}(\gamma'_{n})\frac{\prod_{j=n+1}^{P}\Upsilon_{\{1:M\}}(\gamma_{n},\gamma_{j})}{K_{n}(\gamma'_{1:n-1},\gamma_{n+1:P})},
\end{align*}
where $K_{n}$ is the normalizing constant in the $n^{th}$ sub-iteration
of the block Gibbs sampler. We define $\upsilon_{n}=[-1,-1,0]\otimes[\mathbf{1}_{n}]^{T}$,
where $\mathbf{1}_{n}$ is the $n$-dimensional one vector. If $\gamma'=\upsilon_{P}$,
then for each $j\in\{n+1:P\}$, $\prod_{t=-1}^{0}\prod_{s=1}^{3}(1-1_{\{1:M\}}(\gamma_{j,s})\delta_{t}[\gamma_{j,s}])=1$
as an assignment cannot take -1 or 0 and a positive number at the
same time hence
\begin{align*}
\pi_{n}\left(\upsilon_{P}|\gamma'_{1:n-1},\gamma{}_{n+1:P}\right) & =\prod_{n=1}^{P}\frac{\lambda_{n}^{(I,\xi)}\left(\upsilon_{1}\right)}{K_{n}\left(\upsilon_{n-1},\gamma_{n+1:P}\right)}>0,
\end{align*}
if $\gamma=\upsilon_{P}$ then $1_{\{1:M\}}(\gamma_{j,r})=0$ for
$r\in\{1,2,3\}$, hence
\begin{align*}
\pi_{n}(\gamma'_{n}|\upsilon_{P}) & =\prod_{n=1}^{P}\frac{\lambda_{n}^{(I,\xi)}(\gamma'_{n})}{K_{n}\left(\gamma'_{n},\upsilon_{P-n}\right)}>0.
\end{align*}
Hence, the 2-step probability transition is
\begin{multline*}
\pi^{2}(\gamma'|\gamma)=\sum_{\varsigma\in\Gamma}\pi(\gamma'|\varsigma)\pi(\varsigma|\gamma')>\\
\pi_{n}\left(\gamma'_{n}|\upsilon_{P}\right)\pi_{n}\left(\upsilon_{P}|\gamma'_{1:n-1},\gamma{}_{n+1:P}\right)>0.
\end{multline*}
This condition is sufficient for the block Gibbs sampler to converge
to the target distribution (Proposition 4 in \cite{007}). $\boldsymbol{\enskip\blacksquare}$

\bibliographystyle{IEEEtran}
\bibliography{IEEEabrv,reflib}

\end{document}